\documentclass[10pt, jurnal]{IEEEtran}

\usepackage[applemac]{inputenc}
\usepackage{graphicx,url}
\usepackage{multirow}  
\usepackage{hhline}
\usepackage{graphicx}
\usepackage{subfigure,amssymb,amsmath}
\usepackage{color,soul}
\usepackage[dvipsnames]{xcolor}
\usepackage{bbm}
\usepackage{lipsum}
\usepackage{cfr-lm}
\usepackage{mathtools} %RS 2019-04-30 to enable multi-line equations

\usepackage{balance}

\usepackage{comment}       %RS
\usepackage{dirtytalk}    %RS to enable quotes

\IEEEoverridecommandlockouts
\usepackage[pscoord]{eso-pic}

\title{A Framework for dynamically meeting\\ performance objectives on a service mesh}
\author{\IEEEauthorblockN{Forough Shahab Samani \IEEEauthorrefmark{2} and Rolf Stadler\IEEEauthorrefmark{2}} \protect\\
Dept of Network and Systems Engineering, \protect\\
KTH Royal Institute of Technology, Sweden,\protect\\
Email: \{foro, stadler\}@kth.se}
\begin{document}
\maketitle
\pagenumbering{arabic}

\thispagestyle{plain}
\pagestyle{plain}

\begin{abstract}

We present a framework for achieving end-to-end management objectives for multiple services that concurrently execute on a service mesh. We apply reinforcement learning (RL) techniques to train an agent that periodically performs control actions to reallocate resources. We develop and evaluate the framework using a laboratory testbed where we run information and computing services on a service mesh, supported by the Istio and Kubernetes platforms. We investigate different management objectives that include end-to-end delay bounds on service requests, throughput objectives, cost-related objectives, and service differentiation. We compute the control policies on a simulator rather than on the testbed, which speeds up the training time by orders of magnitude for the scenarios we study. Our proposed framework is novel in that it advocates a top-down approach whereby the management objectives are defined first and then mapped onto the available control actions. It allows us to execute several types of control actions simultaneously. By first learning the system model and the operating region from testbed traces, we can train the agent for different management objectives in parallel. 

\end{abstract}


\section{Introduction}
\label{sec:introduction}

End-to-end performance objectives for a service are difficult to achieve on a shared and virtualized infrastructure, because the service load often changes and service platforms do not offer strict resource isolation. In order to continuously meet performance objectives for a service, such as bounds on end-to-end delays or throughput for service requests, the management system must dynamically re-allocate the resources of the infrastructure. Such control actions can be taken on the physical layer, the virtualization layer, or the service layer, and they include horizontal and vertical scaling of compute resources, function placement, as well as request routing and request dropping.

The service abstraction we consider in this paper is a \emph{directed graph} where the nodes represent processing functions and the links communication channels. This general abstraction covers a variety of services and applications, such as a network slice on a network substrate, a service chain on a softwarized network, a micro-service based application on an IT infrastructure, or a pipeline of machine-learning tasks on a server farm. We choose the service-mesh abstraction in this work and focus on micro-service based applications.

In this paper, we propose a framework for achieving end-to-end management objectives for multiple services that concurrently execute on a service mesh. We apply reinforcement learning (RL) techniques to train an agent that periodically performs control actions to reallocate resources. A management objective in this framework is expressed through the reward function in the RL setup. We develop and evaluate the framework using a laboratory testbed where we run information services on a service mesh, supported by the Istio and Kubernetes platforms \cite{Istio},\cite{K8}. We investigate different management objectives that include end-to-end delay bounds on service requests, throughput objectives, cost-related objectives and service differentiation. 

Training an RL agent in an operational environment (or on a testbed in our case) is generally not feasible due to the long training time, which can extend to weeks for the scenarios we study in this paper, unless the state and action spaces of the agent are very limited. We address this issue by computing the control policies on a simulator rather than on the testbed, which speeds up the learning process by orders of magnitude for the scenarios we study. In our approach, the RL system model is learned from testbed measurements; it is then used to instantiate the simulator, which produces near-optimal control policies for various management objectives, possibly in parallel. The learned policies are then evaluated on the testbed using unseen load patterns (i.e. patterns the agent has not been trained on). 

Compared with other recent research, our proposed framework is unique and novel in the following way. First, it advocates a top-down approach whereby the management objectives are defined first and then mapped onto the available control actions. In all related work we are aware of, resource allocation is investigated as the problem of designing a single control function, such as scaling or request routing, to achieve a particular performance objective in a given scenario. 
Second, our framework allows us to define management objectives for several services and achieve them simultaneously on an infrastructure. Third, several control actions can be executed simultaneously in a single time step of the RL agent. Fourth, our framework is general in the sense that many classes of management objectives as well as a long list of control actions can be supported. In addition, our work is unique in that we learn the operating region for the system model and control functions, which constrains the possible action an agent can take. Moreover, the framework supports different management objectives for the same system model, a useful property that we have not seen discussed anywhere else in the literature.

We make three contributions with this paper. First, we present an RL-based framework that computes near-optimal control policies for end-to-end performance objectives on a service graph. This framework simultaneously supports several services with different performance objectives and several types of control operations. 

Second, as part of this framework, we introduce a simulator component that efficiently produces the policies. While we lose some control effectiveness due to the inaccuracy of the system model, we gain by significantly shortening the training time by two orders of magnitude, which makes the approach suitable in practice. 

Third, we learn the operating region from testbed traces, which constrains the possible action of the agent, and we consider the different settling times of the various control actions, which facilitates meeting the management objectives on the testbed. 

This paper reports on a major extension of a seven-page publication by us published at CNSM 2022\cite{samani2022dynamically}. It contains an improved description of the framework, it discusses an extended set of scenarios, management objectives, and control actions, and it includes a significantly broader related-work section.  

Note that, when developing and presenting our framework, we aim at simplicity, clarity, and rigorous treatment, which helps us focus on the main ideas. For this reason, we choose a small service mesh for our scenarios (which still includes key complexities of larger ones), we consider only three types of control actions, etc. Our plan is to refine and extend the framework in future work, as we lay out in the last section of the paper.

The remainder of this paper is organized as follows. Section \ref{sec:problem_formulation_and_approach} formalizes the problem and details the approach to address it. Section \ref{sec:framework} describes our framework and details of modeling and implementations. Section \ref{sec:testbed_description} details our testbed, the experiments we are conducting, the metrics we are collecting during experiments, and the traces we generate from this data. Sections \ref{sec:usecase} and \ref{sec:implementation and Evaluation set up} discuss the scenarios, learning the control policies and the evaluation results. Section \ref{sec:related_work} surveys related work. Finally, Section \ref{sec:conclusion} presents the conclusion and future work.  

\section{Problem formulation and approach}
\label{sec:problem_formulation_and_approach}

% \begin{itemize}
%     \item The service mesh as a directed graph.
%     \item Management objectives
%     \item The RL formulation
%     \item The role of simulation and of testbed
%     \item Learning the system model
% \end{itemize}

\begin{table}[!htbp]
\centering
\color{black}
\caption{Notation}
\label{tab:notation}
\begin{tabular}{|l|c|}
\hline
\textbf{Concept} & \textbf{Notation} \\\hline \hline
Service index & $i\in\{1,...,m\}$ \\\hline
Node/microservice index & $j\in\{1,...,n\}$ \\\hline
Offered load of service $i$ & $l_i$ \\\hline
Carried load of service $i$ & $l_i^c=l_i(1-b_i)$\\\hline
Utility of service $i$ & $u_i$ \\\hline
Response time objective for service $i$ & $O_i$ \\\hline
Effective response time of service $i$ & $d_i$\\\hline
Routing weight of service $i$ from node $k$  & $p_{ikj}$\\
towards node $j$&\\\hline
Blocking rate for service $i$ & $b_i$\\\hline
Number of CPU cores for node $j$ & $c_j$ \\\hline 
Cost of the service $i$ with $k$ microservices & $g_i=\sum^k_j c_j$\\
\hline 
\end{tabular}
\end{table} 

\begin{figure}[!htbp]
 \centering
 \includegraphics[width=\columnwidth]{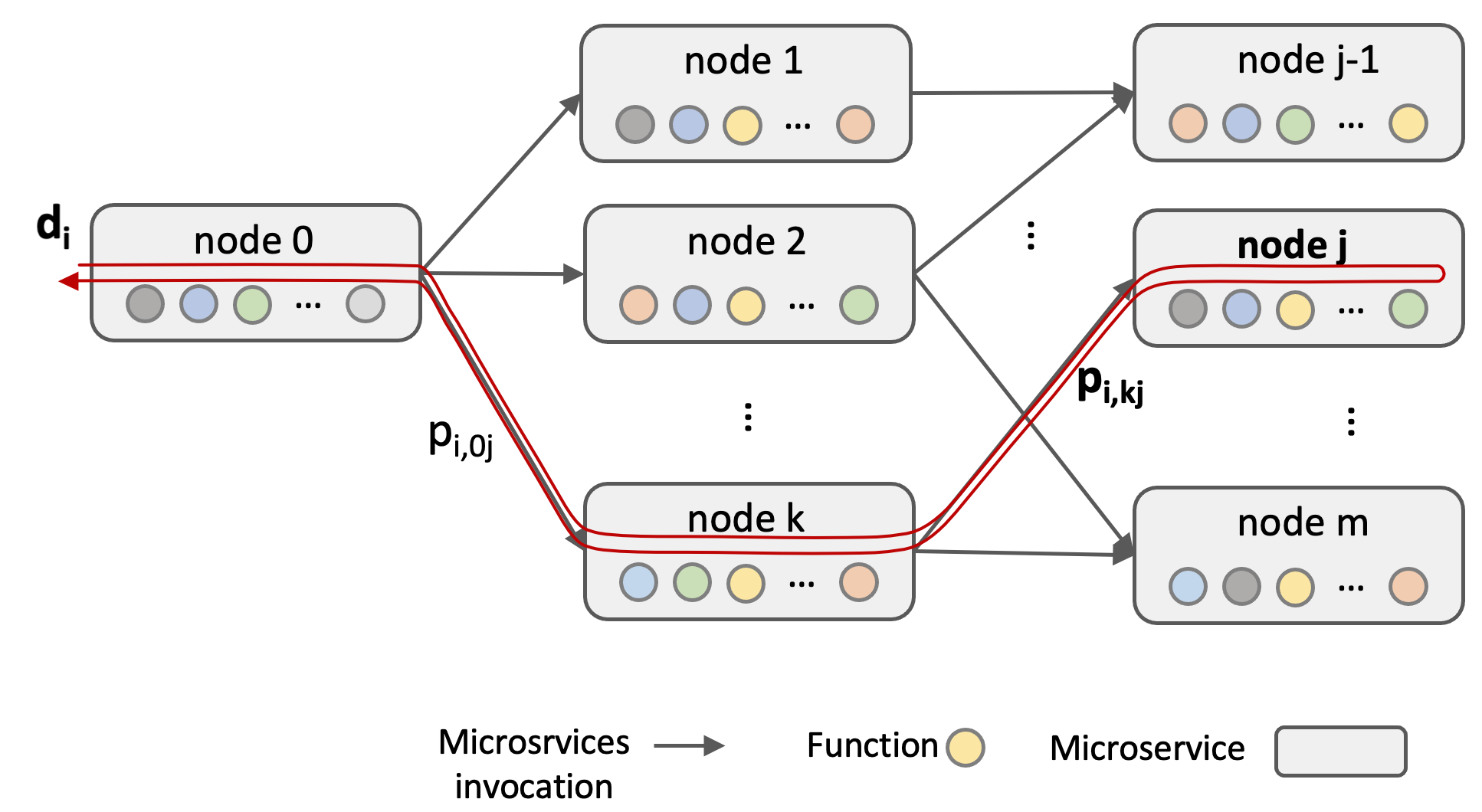}
 \caption{Example of a service mesh whose nodes are microservices and whose links represent microservice invocations. $d_{i}$ is the end-to-end delay of processing a request of service $i$; $p_{ikj}$ denotes the routing weight of microservice $k$ invoking microservice $j$ to process a request for service $i$.}
 \label{fig:service_mesh}
\end{figure}

%================The service mesh as a directed graph.================
\noindent\textbf{Service mesh}. We consider services that are built from microservice components, whereby each component has a dedicated functionality. A service can be understood as a contiguous subgraph on a directed graph which we call the service mesh. A node of the service mesh offers a specific microservice with its functions. A directed link between two nodes represents the invocation of the microservice on the link's head node by the microservice of the tail node. Fig. \ref{fig:service_mesh} gives an example of a service mesh and shows how processing of a service request at node 0 creates a path of service invocations on the service mesh. 

\noindent\textbf{Service metrics}. We associate different metrics with a service $S_i$. In this work, we consider the following service metrics: the (average) end-to-end delay $d_i$ of a service request, which we also call the response time of service $S_i$; the offered load $l_i$ of this service expressed in requests per second; the carried load $l^c_i$ in requests per second (carried load refers to the load that the corresponding microservices receives and reply it, $l^c_i \leq l_i$); the cost of the service $g_i$; and the utility $u_i$ generated by the service.

\noindent\textbf{Management Objectives}. Central to this work are management objectives, which capture the end-to-end performance objectives for the services on a given service mesh. These objectives include client requirements as well as provider priorities. We express a management objective as an optimization objective where a function must be optimized under constraints. The function as well as the constraints contain service metrics as parameters. We study several management objectives in this work. An example is the objective of maximizing the overall throughput of the system (in requests per second) for two services while meeting delay constraints for end-to-end delay $O_i$ for requests of each service.

\begin{equation} \label{eq:mo1}
% \begin{split}
\text{maximize   } \sum_i l_i^c \text{ while } 
d_i<O_i\ \ \ \ i=1,2
% \end{split}
\end{equation}

The service metrics $l^c_i$ and $d_i$ depend on the settings of the control parameters of the underlying resource management system. This means that the decision variables of the optimization problem Eq. (\ref{eq:mo1}) are exactly the control parameters. Note further that environmental variables like the load $l^c_i$ can change over time and that the optimization problem must be solved whenever such a change occurs. 

\noindent\textbf{Control functions}. In order to meet the management objectives, the control parameters of the resource management system are dynamically adjusted. Control parameters we consider in this work concern request routing and are denoted $p_{ikj} \in [0,1]$, which indicates the fraction of requests of service $i$ routed from node $k$ to node $j$. In addition, we consider blocking service requests at the front node and denote the fraction of blocked requests by $b_i \in [0,1]$. Moreover, we consider vertical scaling of the computing resource for each computing node and denote by $c_{j}$ the number of CPU cores allocated to computing node $j$.

\noindent\textbf{Reinforcement learning approach.} We formalize the above problem of optimal control using a Markov Decision Process (MDP) and solve the problem by applying the reinforcement learning approach. Reinforcement learning is a well-established methodology for sequential decision-making \cite{puterman2014markov}\cite{sutton2018reinforcement}. 
We apply it to determine the setting of the control parameters for a networked system that evolves in discrete time steps.

The elements of a reinforcement learning (RL) model include the state space, the action/control space, the system model/transition model, and the reward function. We provide the RL model for the use case explained in Section \ref{sec:usecase}.

%================RL for our problem==============

\noindent \textbf{State space:} The state at time $t$ includes the response time and offered load of the services running on the service mesh: $s_t = (l_{i,t},d_{i,t}|i=1,2)\in \mathcal{S} \subset \mathbb{R}^{2\times 2}$.  

\noindent \textbf{Action/control space:} The action taken at time $t$ is expressed as the control parameter settings at time $t$: $a_t = (b_{i,t}, p_{ijk,t}, c_j|i=1,2, j=1,2) \in \mathcal{A} \subset \mathbb{R}^{2\times 3}$.

\noindent \textbf{System model/transition model}: The model is a function that provides the next state $s_{t+1}$ when a specific action $a_t$ is taken in state $s_t$: 

\begin{equation}
\label{eq:system_model}
    \overbrace{(d_{i,t+1},l_{i,t+1})}^{s_{t+1}} = f(\overbrace{d_{i,t}, l_{i,t}}^{s_t} , \overbrace{b_{i,t} , p_{ijk,t}, c_j}^{a_t} | \ i,j = 1,2)
\end{equation}

We learn the system model through testbed observation and supervised learning. This measurement suggests that the delay at time step $t+1$ does not depend on the delay at time step $t$. Therefore, the system model in Eq. (\ref{eq:system_model}) simplifies to\\
\begin{equation}
\label{eq:system_simple}
    (d_{i,t+1}) = f(l_{i,t} , b_{i,t} , p_{ijk,t}, c_j | \ i,j = 1,2)
\end{equation}

% Since we train the policy on a known load pattern, the system model we must learn can be simplified, as we argue in Section \ref{sec:usecase}.

\noindent \textbf{Reward function}: The reward function $r(s_t,a_t)$ computes a numerical reward if action $a_t$ is taken in the state $s_t$. In our context, the reward function expresses the management objective in the RL model. 
The reward function that expresses the example management objective given in $(\ref{eq:mo1})$ can be written as 

$$r(s_t,a_t)= l_1^c \times r_1(s_t,a_t) + l_2^c \times r_2(s_t,a_t)$$ 

\noindent whereby $r_1$ and $r_2$ are functions based on $tanh$ (see Section \ref{sec:usecase}). The reward functions of all management objectives studied in this work are detailed in Section \ref{sec:usecase}.

\section{Solution Framework}
\label{sec:framework}

% - Introduction:
%   1) Start with the overall objective and at the end, we have the control policy on the testbed
%   2) We use the RL approach, use a simulator and learn the system model explicitly
% - Motivation why we use the simulator:
%   1) Long training time for RL agents on the real system. 
%   2) Edge cases, where we cannot take any action
%   3) Training RL agents with different management objectives on the simulator that is trained only once. 
% - Description of the method and the procedure, here we describe the figures and also how components talk to each other.
Starting from an overall management objective and intent, the goal is to find an optimal (or close-to-optimal) policy that  provides the control settings for a given state of the target system (in this work target system and testbed are used interchangeably). Moreover, these settings should adapt as the system state and the environment evolve. In this section, we present the solution framework for this engineering task.

Compared with the recent literature, our framework has the following distinguishing features. (See Section \ref{sec:related_work} for more details on related work).
First, the control policies are determined by end-to-end management objectives, which can be of varying complexity. They are defined using a set of metrics that can be observed on the target system. 
Second, the framework includes the target system on which the RL agent acts, as well as a simulator where effective policies are computed in an efficient way. 
The key reason for including a simulator is the long and often unfeasible training time for an RL agent operating on the target system. Agent training includes a series of interaction steps, each of which is composed of the settling time after a control action followed by the observation time of the new system state. For the scenarios considered in this work, an interaction step takes seconds to minutes on the target system, and thousands of such steps are typically needed to fully train an agent. In the simulation environment, a step can be executed within milliseconds, which speeds up the training time by several orders of magnitude. In addition, the simulator allows us to study the robustness or generalization properties of a learned policy, for instance by observing the effect of changes in the load pattern, the system configuration, or environmental parameters, while keeping the policy constant.   
Third, we explicitly learn the system model from monitoring the target system. This opens up the possibility to learn different policies for different management objectives for the same system model.
Finally, the framework allows any number of control actions (e.g., scaling, routing, and blocking), which can be executed in sequence or concurrently and which allows for a rich set of control actions in support of meeting the management objectives. In comparison, many recent works focus on studying a single type of control action for a given performance objective.

\begin{figure}[!htbp]
 \centering
 \includegraphics[width=\columnwidth]{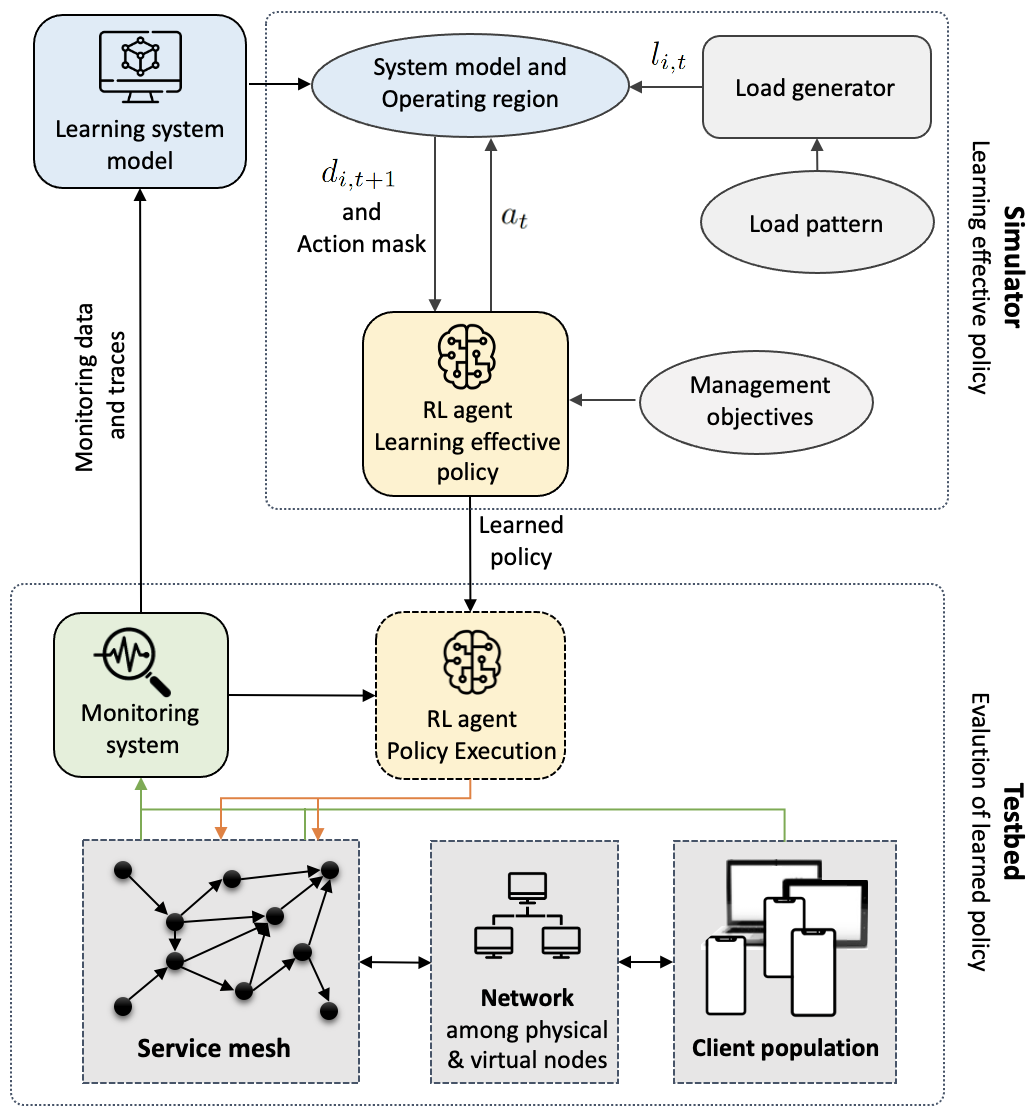}
 \caption{Structure and components of solution framework.}
 \label{fig:solution_framework}
\end{figure}

Figure \ref{fig:solution_framework} shows the solution framework of our method including processes and the data flows between them. To find optimal control actions in support of an end-to-end management objective, we perform the following tasks in a sequential fashion.   

\noindent\textbf{$1$ Describe use case, management objective(s), control actions.} We describe the use case by giving details about the infrastructure, the services, the stakeholders, etc. Then, we formulate the management objective (of the service provider) and the control functions in the infrastructure that are available to achieve the management objective. (See Section \ref{sec:problem_formulation_and_approach}.)

\noindent\textbf{$2$ Develop the RL model including the reward function.} Here we formalize the use case and the management objective using the reinforcement learning approach. This includes defining the key elements of the RL model, namely, states, actions, and a reward function that expresses the management objective. (See Section \ref{sec:problem_formulation_and_approach}.)

\noindent\textbf{$3$ Define a scenario and run it on the target system.} Within the context of the use case, we define several scenarios which approximate operational conditions and run them on the target system. In this paper, we report on four scenarios. (See section \ref{sec:usecase}.) A scenario description contains the characterization of the offered load, the service configuration, and the management objective. We set up a scenario on the target system, run it, and collect data traces through periodic monitoring of the system state and service metrics. A data item has the format of $(s_t,a_t,s_{t+1})$, whereby $s_t$ describes the system state at time $t$, $a_t$ the value of the control parameters at time $t$, and $s_{t+1}$ the new state of the system after the control parameters have been applied.  

\begin{figure}[!htbp]
 \centering
 \includegraphics[width=\columnwidth]{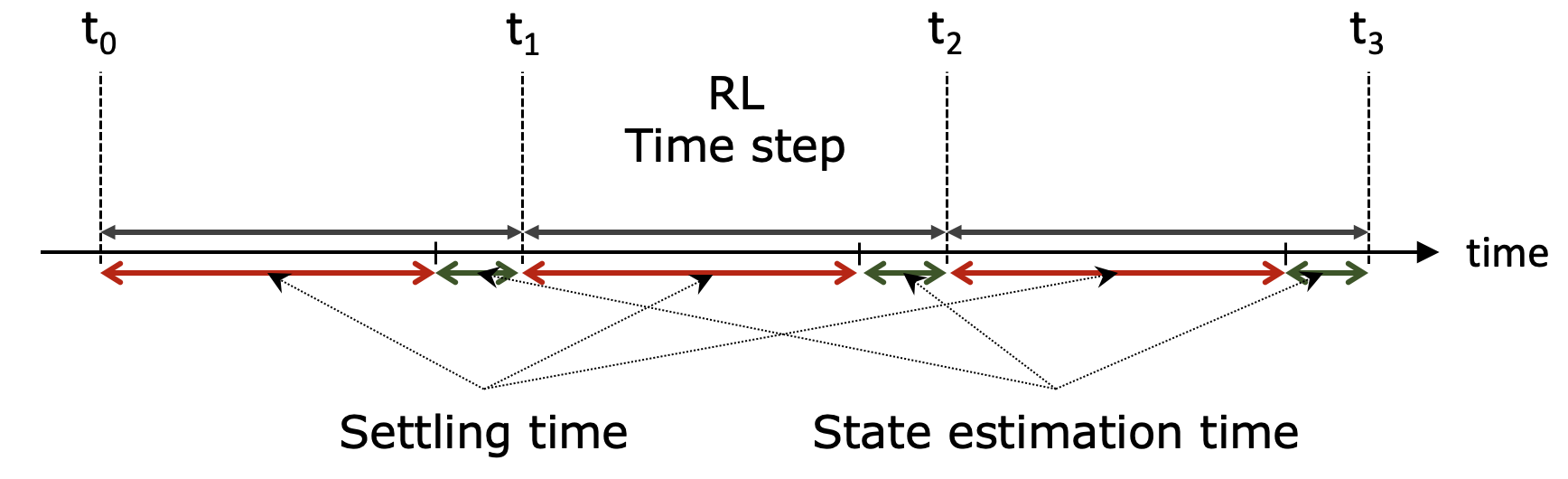}
 \caption{Time step of the RL model on the target system. It is composed of the settling time and time for state estimation time. }
 \label{fig:time_step}
\end{figure}

In the RL model, time evolves in discrete steps. We choose the duration of a time step on the target system in such a way that it includes the settling time of the control action and the time it takes to estimate the new system state. (By settling time we understand the time it takes to reach an equilibrium state after taking a control action \cite{franklin2002feedback}). Figure $\ref{fig:time_step}$ shows the time steps of the reinforcement learning model on the target system. The settling time depends on the type of control function. In the experiments for this work, the settling time for routing action is typically about one second, and the settling time for scaling is generally about $30$ seconds. 

\noindent\textbf{$4$ Estimate the system mode and the operating region.} The system model is a function that maps the system state to a new state after applying a specific control function \cite{bertsekas2019reinforcement}. In this work, the system model is given by Eq. $(\ref{eq:system_simple})$. We estimate the system model through supervised learning using random forest regression $\cite{BR01}\cite{RF_sklearn}$ from the collected data traces. 

\begin{figure}[!htbp]
 \centering
 \includegraphics[width=0.6\columnwidth]{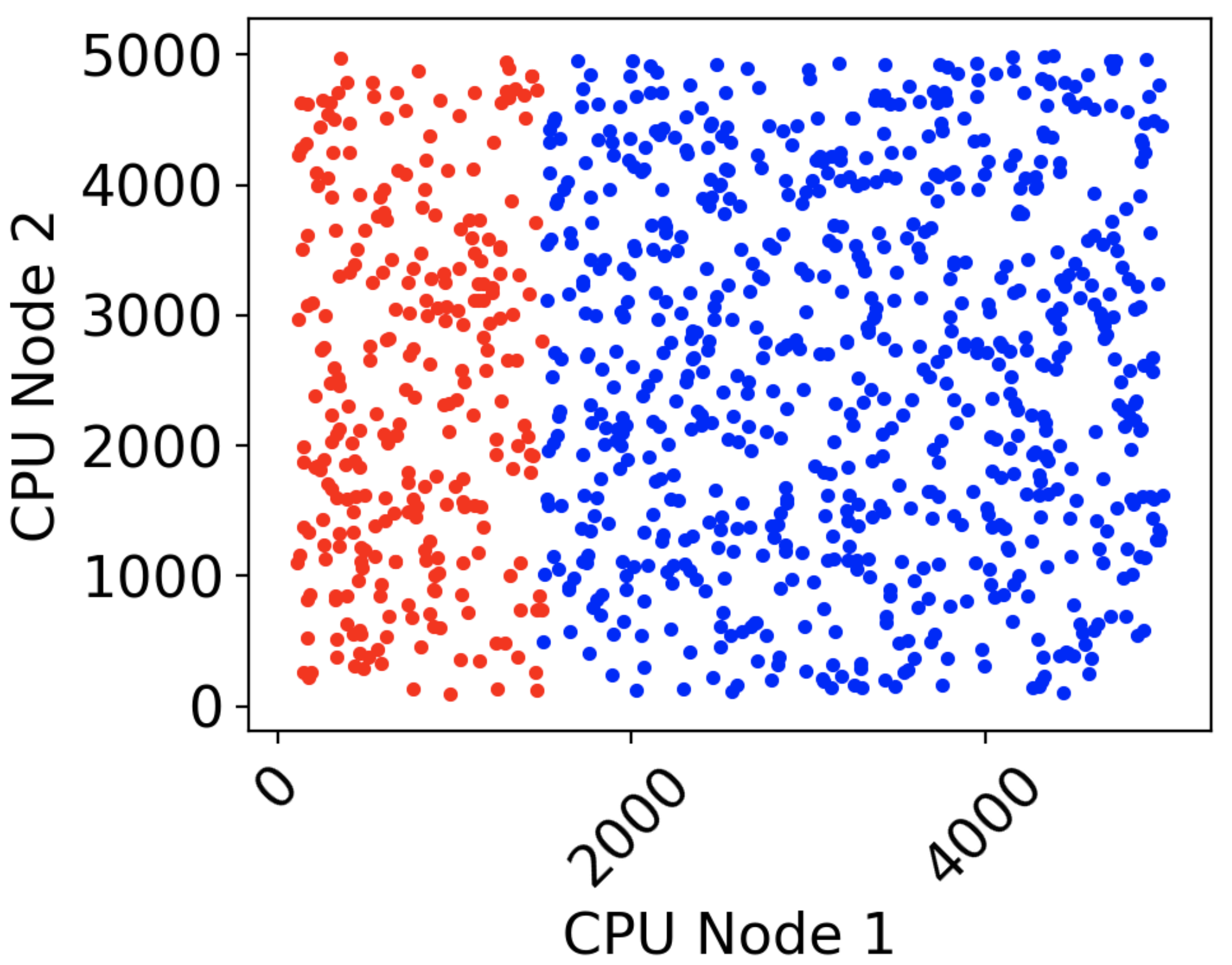}
 \caption{Cut of the operating region (blue dots) for the services on the target system for fixed offered load, and fixed settings of routing and blocking parameters (i.e., $l_1=0,l_2=20,l_3=5,p_{11}=p_{21}=p_{31}=0.2,b_1=b_2=b_3=0$).}
 \label{fig:op}
\end{figure}

To achieve predictable and stable service quality, we want the system to be stable and perform within its operating region \cite{hellerstein2004feedback}. We say that a tuple consisting of the system state $s$ and the setting of control parameters $a$ is within the operating region $O_p$ of the system if the state is stationary. For our use case, the system state is stationary if the mean response time remains constant and its variance is small. (In this work, we consider the variance to be small if it is smaller than $50\%$ of the mean.) Formally,

\begin{equation}
\begin{split}
\label{eq:op}
    (s,a)\in O_p \subset \mathcal{S}\times\mathcal{A} \iff \text{ mean of } s \text{ is constant} \\ \text{ and its variance is small.}
\end{split}
\end{equation}
Figure \ref{fig:op} illustrates an example of the operating region for the services running on the target. 

\noindent\textbf{$5$ Train and evaluate the RL agent on the simulator.} The upper part of Figure \ref{fig:solution_framework} shows the structure and the main components of the simulator. The system model receives as input the offered load from the load generator and the actions from the RL agent at each time step, and it predicts the response time of the services for the next time step. The RL agent computes the reward and updates the control policy.
Note that the action that the RL agent takes is constrained by the operating region which is expressed by the action mask in Figure $\ref{fig:solution_framework}$ \cite{huang2020closer}\cite{maskable_PPO}.

\noindent\textbf{$6$ Evaluate the RL agent on the target system.} We evaluate the learned policy with respect to robustness and generality and with respect to its effectiveness on the target system. To assess the generality of the policy, we first evaluate the learned policy on the simulator using a different load pattern (i.e., an operational load pattern) than the agent has been trained on. This informs us about the generalization error of the learned policy. Second, and more important, we evaluate the learned policy on the target system for the load pattern used to train the agent as well as the operational load pattern. The gap in effectiveness between the simulation results and the results from the target system is determined by the accuracy of the system model. Finally, we assess the robustness of the policy by evaluating the learned policy on the target system using a different load pattern than the agent has been trained on.

\section{Target system: The KTH testbed}
\label{sec:testbed_description}
% \begin{itemize}
%     \item Hardware and software stack
%     \item The services 
%     \item Generating the load
% \end{itemize}

% Figure \ref{fig:testbed} outlines our laboratory testbed at KTH. 

\begin{figure}[!htbp]
 \centering
 \includegraphics[width=0.9\columnwidth]{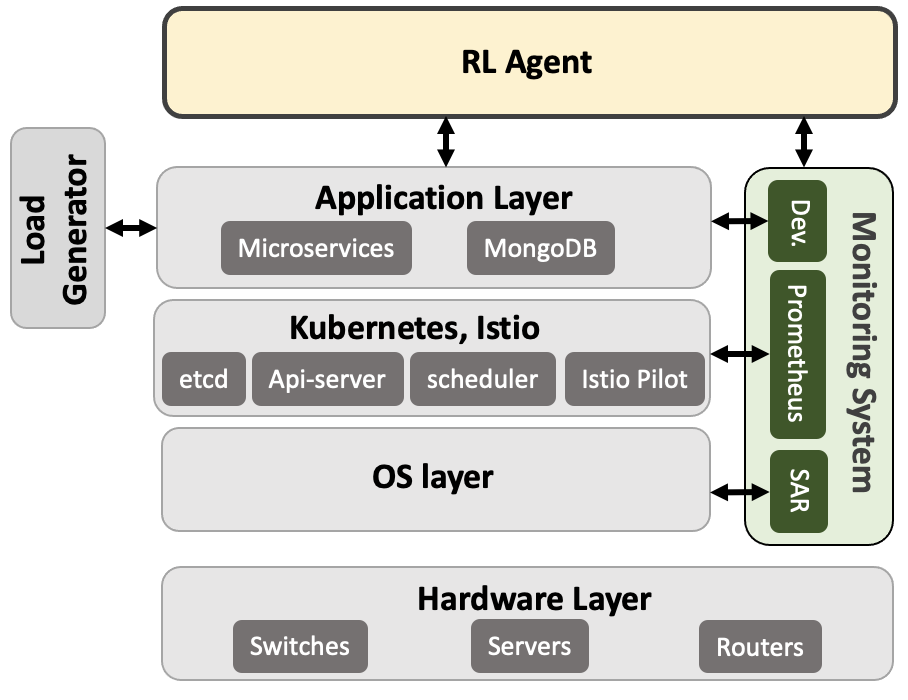}
 \caption{Software stack of the testbed at KTH.}
 \label{fig:swstack}
\end{figure}

\noindent\textbf{Hardware and orchestration layers.} Our testbed at KTH includes a server cluster connected through a Gigabit Ethernet switch. The cluster contains nine Dell PowerEdge R$715$ $2$U servers, each with $64$ GB RAM, two $12$-core AMD Opteron processors, a $500$ GB hard disk, and four $1$ Gb network interfaces. The tenth machine is a Dell PowerEdge R$630$ $2$U with $256$ GB RAM, two $12$-core Intel Xeon E$5$-$2680$ processors, two $1.2$ TB hard disks, and twelve $1$ Gb network interfaces. Figure $\ref{fig:swstack}$ shows the software stack. 
All machines run Ubuntu Server $18.04.6$ $64$ bits and their clocks are synchronized through NTP \cite{NTP}. The orchestration layer and the service mesh are realized using Kubernetes (K8) \cite{K8} and Istio \cite{Istio}.

\begin{figure}[!htbp]
 \centering
 \includegraphics[width=0.8\columnwidth]{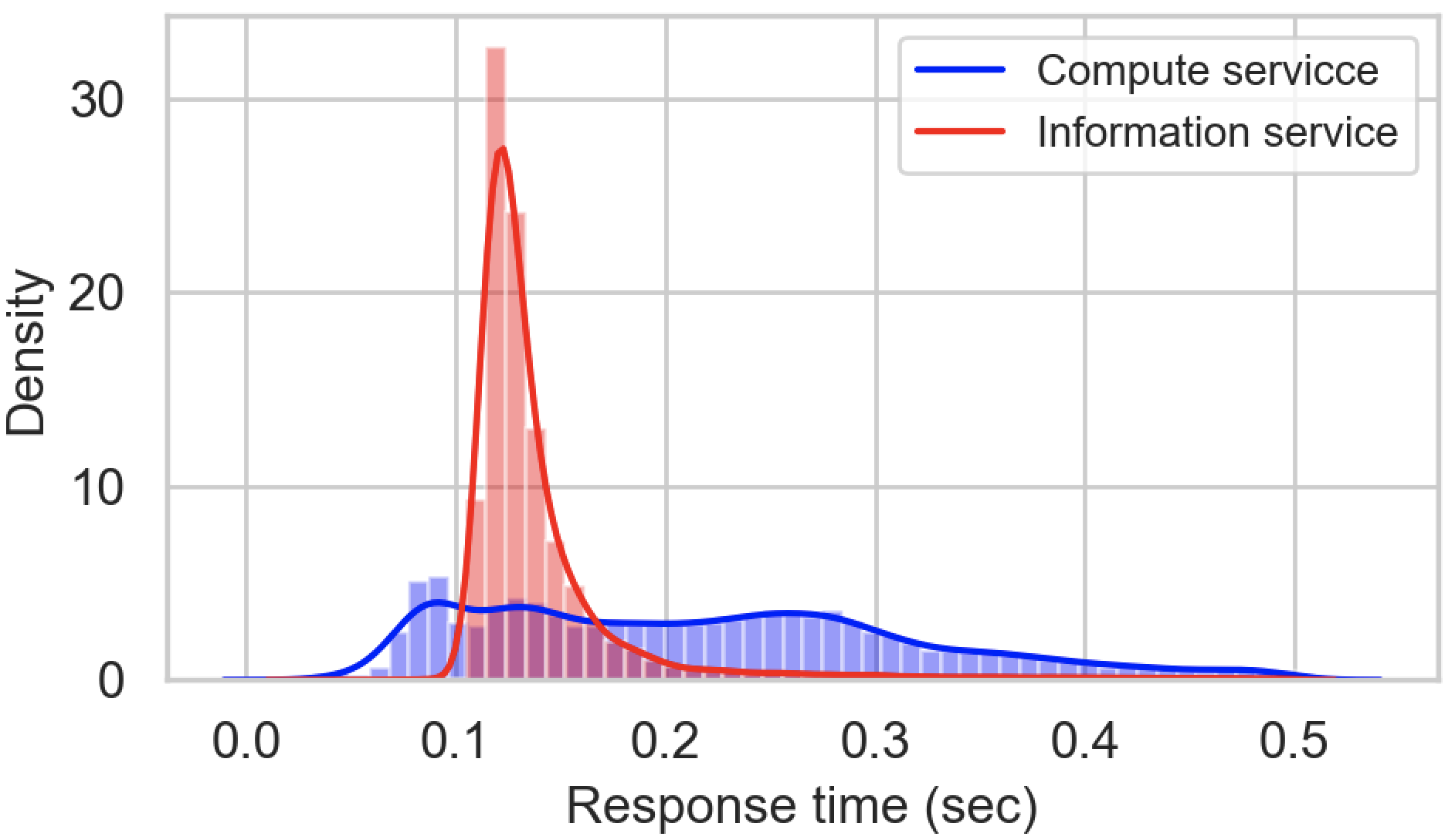}
 \caption{Response time distribution of an information service and a compute service on the testbed.}
 \label{fig:services_response_dist}
\end{figure}

\noindent\textbf{Services.} Using Istio to handle and route the service requests, we have implemented three services. Two of them are information services related to an electronic bookstore and people registry, respectively. A service request contains an object identifier and the response contains information about this object. The services differ in terms of the structure of the objects and the size of the databases. The third service is a compute service that multiplies two random matrices of the size $2\,000\times1\,000$ and $\,1000\times 2\,000$. Figure $\ref{fig:services_response_dist}$ illustrates the response time distribution of one of the information services and the compute service.    

\begin{figure}[!htbp]
 \centering
 \includegraphics[width=\columnwidth]{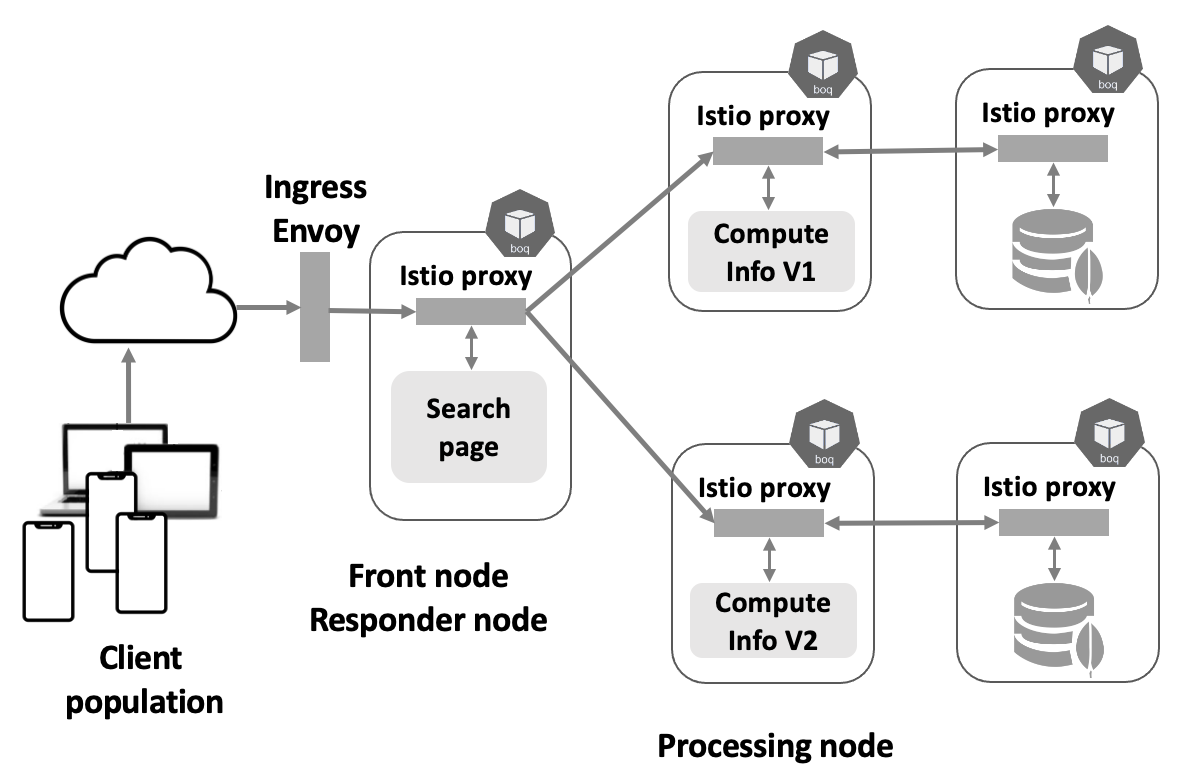}
 \caption{The service mesh with five microservices that implement the three services.}
 \label{fig:application}
\end{figure}

Figure \ref{fig:application} shows the implementation of the services on the testbed. All nodes are implemented in Python \cite{python} using Flask \cite{flask}. The front node provides the web user interface which receives the service requests from the clients and sends responses to these requests. Based on the service type, it invokes an appropriate function on a connected node. Both nodes to the right of the figure include the databases for both information services. Therefore, each information service can be accessed through the upper or lower path in the figure. The compute service can be accessed through either of the two processing nodes. Each of the five nodes in Figure \ref{fig:application} is implemented as a Kubernetes pod \cite{pods}. 

Figure $\ref{fig:services_response_dist}$ shows the distribution of the response times of service requests for one of the information services and the compute service. To produce this figure, we run services on the system for different values of configuration parameters (e.g., routing and scaling) and under varying offered load following the random load pattern explained below.

\noindent\textbf{Generating service load.} We have implemented a load generator in order to train the RL agent or emulate a client population. The load generator is driven by a stochastic process that creates a stream of service requests. We realize two load patterns with this generator.\\ 
The \textit{random load pattern} is used for training. It produces a stream of requests at the rate of $l_i(t)\sim\mathbbm{U}_{\{5,10,15,20\}}$ requests/second for information services and $l_i(t)\sim\mathbbm{U}_{\{1,2,3,4,5\}}$ for the compute service. At every time step, it changes to a value drawn uniformly at random from ${\{5,10,15,20\}}$ for information services and from ${\{1,2,3,4,5\}}$ for the compute service. A time step is $5$ seconds on the testbed for Scenarios $1-3$ and one minute for Scenario $4$. (See Section \ref{sec:usecase}) \\
The \textit{sinusoidal load pattern} approximates the behavior of the clients in an operational setting. It generates a stream of requests at the rate of $l_i(t) = 12.5 + 7.5 \times \sin (\frac{2\pi}{T} t + \phi)$ requests/second for information service $i$ and $l_i(t) = 3 + 2 \times \sin (\frac{2\pi}{T} t + \phi)$ for compute service $i$.  

\section{Scenarios}
\label{sec:usecase}

% In this paper, we model the presented use case as a directed graph shown in Figure \ref{fig:use_case_graph}. In this figure, each box shows a microservice, and circles within these boxes show the functions used to define the services.  

% \begin{figure}[!htbp]
%  \centering
%  \includegraphics[scale=0.4]{images/use case and scenarios/usecase_graph.png}
%  \caption{Directed graph of the application use case used in this paper. }
%  \label{fig:use_case_graph}
% \end{figure}

% For the given service mesh, we define four management objectives formulated within four scenarios.

\begin{figure}[!htbp]
 \centering
 \includegraphics[width=0.8\columnwidth]{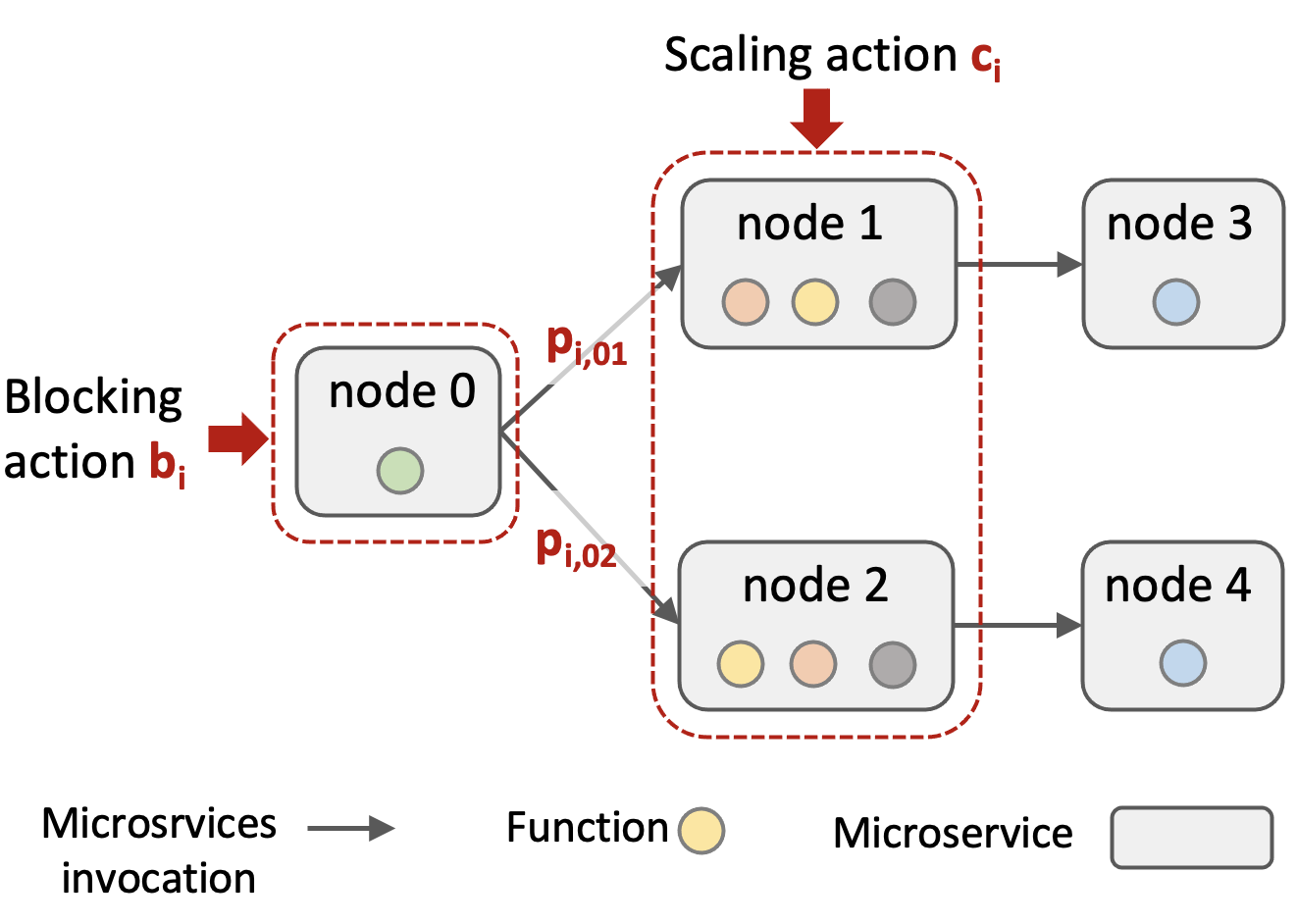}
 \caption{Directed graph of the application use case used in this paper. }
 \label{fig:use_case_graph}
\end{figure}

Following the use case that includes an IT infrastructure, an Orchestration layer with Kubernetes and Istio, and the microservice application described in Section \ref{sec:testbed_description}, we describe four scenarios that will be used to evaluate our solution framework. Figure \ref{fig:use_case_graph} shows the directed graph of services presented in this work. 

\noindent\textbf{Scenario $1$.} In this scenario, we run two information services, which we call information service $1$ and information service $2$. The management objective is that the overall carried load $l_i^c$ is maximized while the response time of each service $d_i$ is bounded by $O_i$:

\begin{equation} \label{eq:mo1_1}
% \begin{split}
\text{maximize   } \sum_i l_i^c \text{ while } 
d_i<O_i \ \ \ \ i=1,2
% \end{split}
\end{equation}

\begin{figure}[!htbp] 
  \centering
  \subfigure[$r_1(s_t,a_t)$]{\label{fig:sc1_reward_2d_s1}
     \includegraphics[width=0.45\columnwidth]{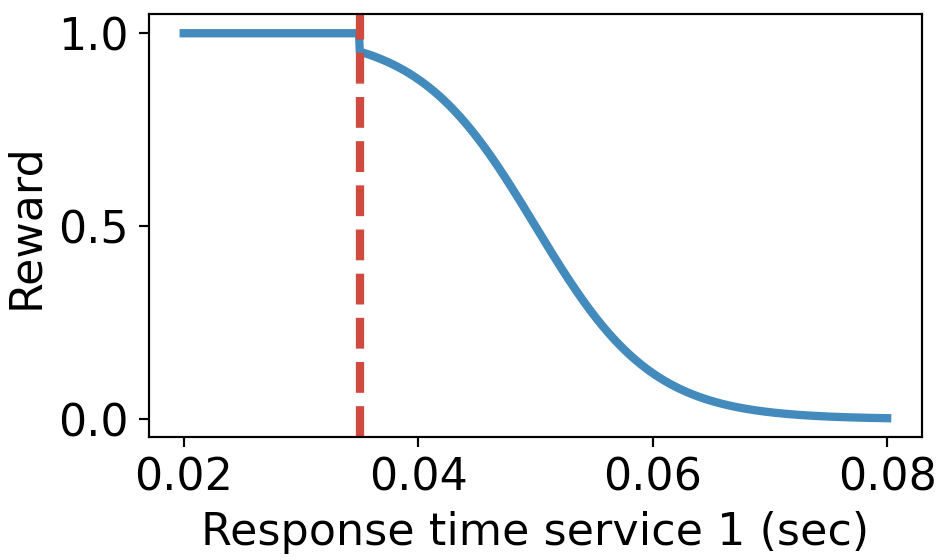}}
  \subfigure[$r_2(s_t,a_t)$]{\label{fig:sc1_reward_2d_s2}
     \includegraphics[width=0.45\columnwidth]{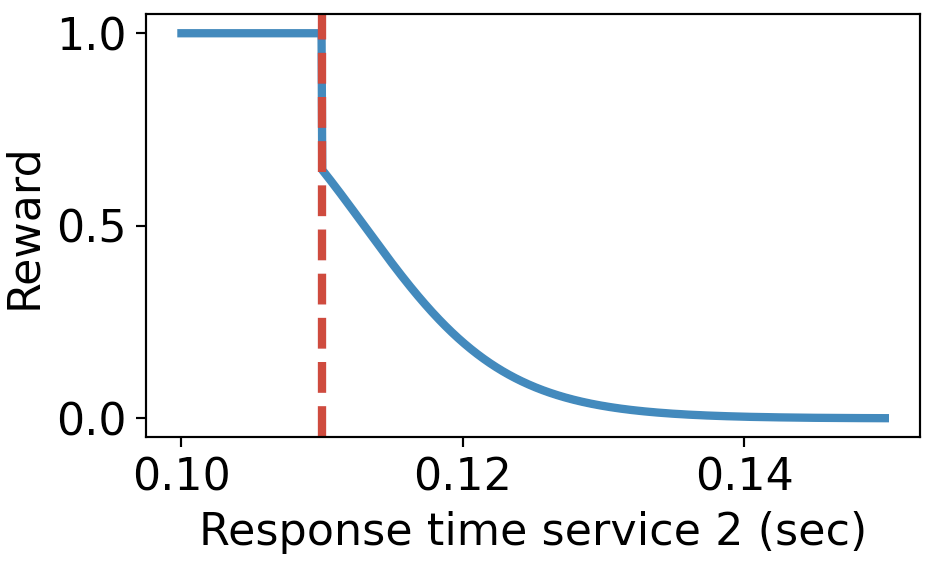}} 

  \caption{The components of the reward functions for management objectives for scenarios 1, 2, and 3.}
  \label{fig:sc1_rewards_2d}
\end{figure}

The RL agent attempts to maximize the joint throughput while enforcing constraints on the response times under changing load. The reward function, which expresses the management objective, is $$r(s_t,a_t)= l_1^c \times r_1(s_t,a_t) + l_2^c \times r_2(s_t,a_t)$$.

\noindent whereby $r_1$ and $r_2$, depicted in Figure $\ref{fig:sc1_rewards_2d}$ are based on the $tanh$ function, capturing the reward of the average response time during time step $t$ for both information services. Both functions $r_1$ and $r_2$, express soft constraints with respect to management objective Eq. $(\ref{eq:mo1_1})$. (If they would express hard constraints, they would be step functions) Soft constraints reflect the fact that a small violation of the response time bound $O_i$ is more valuable than a large violation.   

The control actions for this scenario are routing weights and the blocking rates for both information services (see Figure \ref{fig:use_case_graph}).  

%Use extended figure

%%%%%%%%%%%%%%%%%%%%%%%%%%%%%%%%%%%%%%%%%%%%%%%%%%%%%%%%%%%%%%%%%%%%%%%%%
\noindent\textbf{Scenario $2$.} Similar to Scenario $1$, in this scenario, we run information services $1$ and $2$. The management objective is that the sum of service utilities $u_i$ is maximized while the response time of each service $d_i$ is bounded by $O_i$:

\begin{equation} \label{eq:mo2}
% \begin{split}
\text{maximize   } \sum_i u_i \text{ while } d_i<O_i\ \ \ \ i=1,2
% \end{split}
\end{equation}

The RL agent attempts to maximize the joint service utilities while enforcing constraints on the response times under changing load. The utility functions are defined using the throughput of the services. The reward function, which expresses the management objective, is $$r(s_t,a_t)= \overbrace{l_1^c}^{u_1} \times r_1(s_t,a_t) + \overbrace{5 \times l_2^c}^{u_2} \times r_2(s_t,a_t)$$.

where $r_1$ and $r_2$ are defined above. Similar Control actions to scenario $1$ are used in this scenario. 

%%%%%%%%%%%%%%%%%%%%%%%%%%%%%%%%%%%%%%%%%%%%%%%%%%%%%%%%%%%%
\noindent\textbf{Scenario $3$.} Similar to previous Scenarios, in this scenario, we run information services $1$ and $2$. The management objective is that the carried load $l_i^c$ of service $i$ is maximized, while service $k$ is prevented from starving (i.e., by specifying a lower threshold $l_{min}$ for service $k$) as well as the response time of service $i$ is upper bounded by $O_i$:

\begin{equation} \label{eq:mo3}
\begin{split}
\text{maximize   } l_i^c \text{ while }d_i<O_i \\ \text{ and } l_k^c>l_{min} \ \ \ i\neq k, i=1,2
\end{split}
\end{equation}

The RL agent attempts to maximize the throughput of service $i$ while enforcing constraints on the throughput of service $k$ and response time of service $i$ under changing load. The reward function, which expresses the management objective, is 
$$r(s_t,a_t)= l_2^c \times\bigl(r_3(s_t,a_t) + r_2(s_t,a_t)\bigl)$$
where $r_3$ is a function based on $tanh$ to capture the reward of the carried load during time step $t$ for service $k$. Similar Control actions to scenario $1$ are used in this scenario. 

%%%%%%%%%%%%%%%%%%%%%%%%%%%%%%%%%%%%%%%%%%%%%%%%%%%%%%%%%
\noindent\textbf{Scenario 4.} In this scenario, we run one Information service $2$ and one compute service, that we call it service $3$. The management objective is that the overall cost $g_i$ is minimized while the response time of each service $d_i$ is bounded by $O_i$:

\begin{equation} \label{eq:mo4}
% \begin{split}
\text{minimize   } \sum_j g_i \text{ while } d_i<O_i\ \ \ \ i=1,2
% \end{split}
\end{equation}

The RL agent attempts to minimize the joint cost while enforcing constraints on the response times under changing load. The reward function, which expresses the management objective, is $$r(s_t,a_t)= C(s_t,a_t)\times\bigl(r_2(s_t,a_t) + r_4(s_t,a_t)\bigl)$$
where $r_4$ is a function based on $tanh$, capturing the reward of the average response time during time step $t$ for compute service, and $C(s_t,a_t)$ a function that captures the cost of CPU utilization at time step $t$ for each service. This function linearly decreases with the increase of the CPU utilization of the microservices of the service. In this scenario, we use scaling and routing as the control functions.

Note that our use case allows for many more management objectives than included in the above scenarios. For instance, a management objective can include percentiles of the response times or the variance of the response times. Further, the management objective could focus on minimizing blocked requests or maximizing throughput while keeping costs under a certain level, etc.

\section{Evaluation set up}
\label{sec:implementation and Evaluation set up}

\noindent\textbf{Estimating the system model and operating region.}

In a first run on the testbed, we estimate the system model Eq. ($\ref{eq:system_simple}$) and the operating region Eq. ($\ref{eq:op}$) for Scenarios $1,2 \text{ and }3$. (Note that while these three scenarios have different management objectives, they rely on the same system model.) We run information services $1$ and $2$, for which the load is produced independently by the random load pattern described in Section \ref{sec:testbed_description}. This experiment extends over $5$ days, during which we collect some $45\,000$ samples, one sample every $5$ seconds. The structure of a sample is described in Section \ref{sec:framework}.

In a second run, we estimate the system model and the operating region for Scenario $4$. We run the information service $2$ and the compute service, for which the load is produced independently. The duration of this experiment is  one week, during which we collect some $36\,000$ samples, one sample every $5$ seconds. The longer duration of this run is due to the settling time of the scaling action, which is around $60$ seconds.

From the collected data, we estimate the system model using random forest regression. The applied regressor includes $120$ estimators. To obtain the system model, we estimate the average response time for a given load $s$ and a given setting of the control parameters $a$. To obtain the operating region, we estimate in addition the variance of the response time for each pair $(s,a)$ (see Eq. $\ref{eq:op}$).   

\noindent\textbf{Training the RL agent.} The RL agent is implemented using the PPO algorithm from the Stables Baselines$3$ library \cite{schulman2017proximal}\cite{baselines3}. We configure the actor and critic neural networks to have two hidden layers of size $64$. We set the batch size for training to $64$, the distance between updates to $1\,024$ steps, the learning rate $\alpha = 0.001$, and the discount factor $\gamma = 0$. The number of iterations for the agent to converge falls between $2\,000$ and $5\,000$ depending on the scenario.

When training starts, the agent is initialized with a random policy. During a simulation run, we track the learning process by collecting the reward and the Normalized Reward ($NR$) values (see below). We evaluate the learned policy for $100$ time steps at regular intervals during training and compute their $ANR$ values and produce a learning curve. The simulation terminates when the values on the learning curve have sufficiently converged.

Note that while three scenarios share the same system model, we must train the RL agent for each scenario separately, because each scenario has a different management objective and hence a different control policy. However, these policies can be computed in parallel by running several instances of RL agents and simulators concurrently.   

\noindent\textbf{Estimating the optimal policy.} For a given state $s_t$, we compute the rewards of all possible actions $a_t$ using the simulation model and the reward function. The maximum reward over all possible actions in the operating region is the \emph{optimal reward}, and the action $a'_t$ giving this reward determines the \emph{optimal policy}. 

\noindent\textbf{Evaluation metric for learned policy.} Given a state and an action $(s_t,a_t)$, we define the Normalized Reward $NR$ as the ratio between the  reward obtained by the agent and the optimal reward. 

\begin{equation}
\label{eq:metric}
    NR(s_t,a_t) = \frac{\text{agent reward}(s_t,a_t)}{\text{optimal reward}(s_t,a'_t)}
\end{equation}

For a time interval [1, ..., $T$] we define the Average Normalized Reward (ANR) as  
\begin{equation}
\label{eq:metric}
    ANR = \frac{1}{T}\sum_{t=1}^T NR(s_t,a_t)
\end{equation}

The values of $NR$ and $ANR$ are between $0$ and $1$. $1$ means the management objective is met (throughout the time interval). 

\noindent\textbf{Evaluation metric for the system model.}
To evaluate the system model's prediction accuracy we use Normalized Mean Absolute Error (NMAE) which is defined as $
\frac{1}{\bar{y}}(\frac{1}{m}\sum\limits_{i=1}^m|y_i-\hat{y_i}|)$, where $\hat{y_i}$ is the mean value of predicted values for the target variables, $\bar{y}$ is the average value of the target variable, and m is the number of samples. NMAE is an intuitive and useful measure in the application domain.

%%%%%%%%%%%%%%%%%%%%%%%%%%%%%%%%%%%%%%%%%%%%%%%%%%%%%%%%%%%%%%%%%%%%%
%%%%%%%%%%%%%%%%Evaluation of the scenarios%%%%%%%%%%%%%%%%%%%%%%%%%%
%%%%%%%%%%%%%%%%%%%%%%%%%%%%%%%%%%%%%%%%%%%%%%%%%%%%%%%%%%%%%%%%%%%%%
\section{Evaluation results and discussion}

\begin{figure}[!htbp]
 \centering
 \includegraphics[width=0.8\columnwidth]{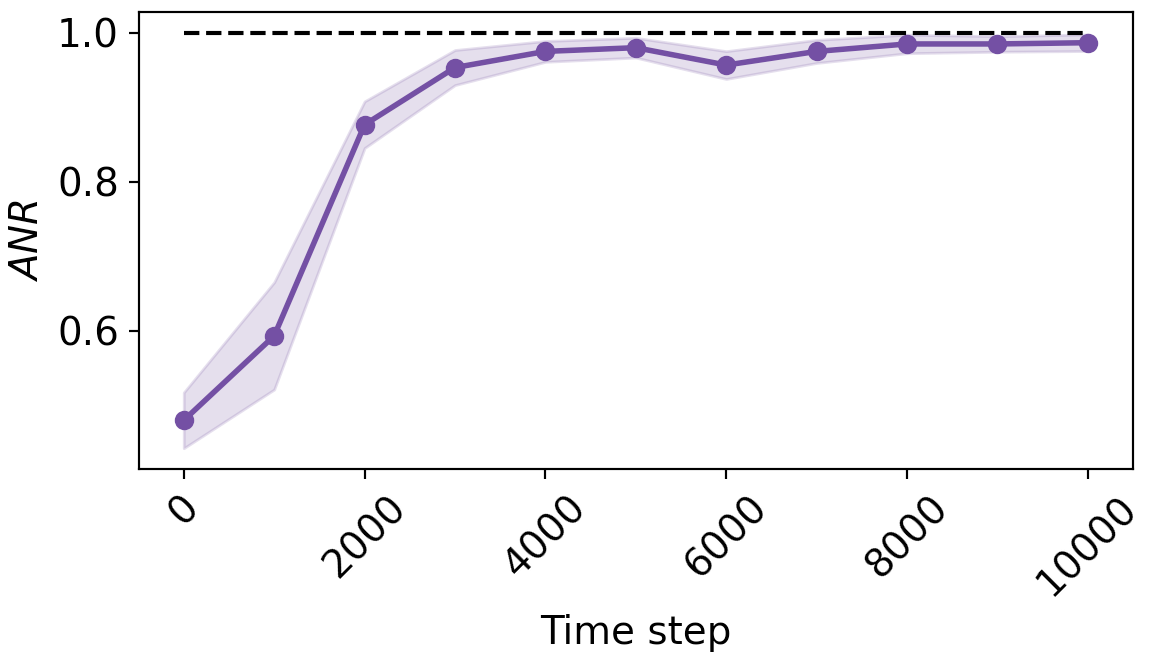}
 \caption{The learning curve of the RL agent for performance metric $ANR$ on the \textbf{simulator} running Scenario $1$. Each point is the average $ANR$ of $100$ steps; the colored area is the $95\%$ confidence interval of the mean $ANR$ values. }
 \label{fig:learning_curve}
\end{figure}

%%%%%%%%%%%%%%%%%%%Evaluation results of Scenario 1%%%%%%%%%%%%%%%%%%
\noindent\textbf{Scenario $1$: agent training.} Having learned the system model, we run Scenario $1$ on the simulator and train the RL agent in order to learn the control policy that meets $MO1$ Eq. $(\ref{eq:mo1_1})$. After every $1000$ simulation steps, we evaluate the learned policy for $100$ time steps and compute the $ANR$ value. We produce the learning curve with the $ANR$ values, which is shown in Figure \ref{fig:learning_curve}. We conclude from the curve that the agent learns an increasingly effective policy. The $ANR$ value approaches $0.98$ after some $4\,000$ steps.

\begin{figure}[!htbp] 
  \centering
  \subfigure[Offered and carried load]{\label{fig:sc1_sim_seen_ls}
     \includegraphics[width=\columnwidth]{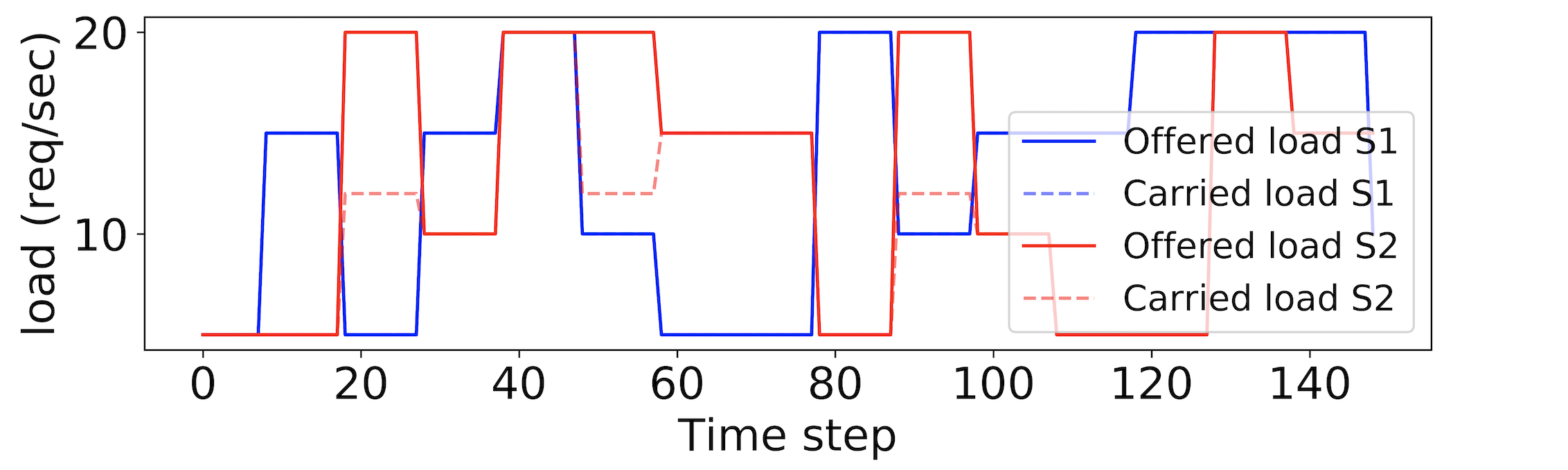}} \hspace{0.15cm}
  \subfigure[Performance of the RL agent]{\label{fig:sc1_sim_seen_nr}
     \includegraphics[width=\columnwidth]{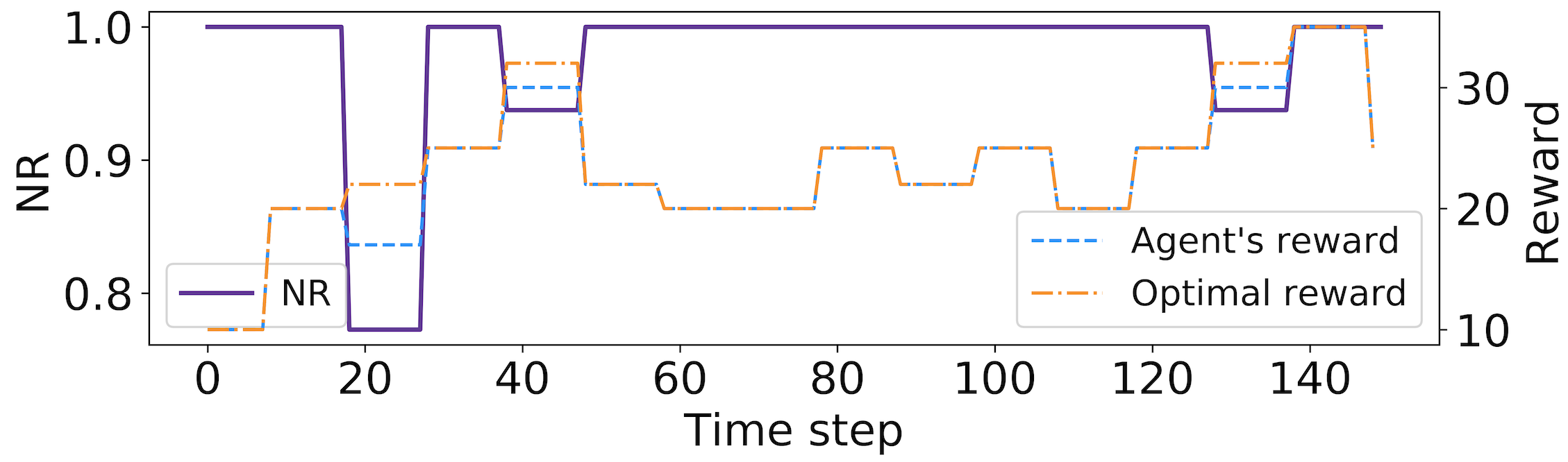}}
     
  \caption{\textbf{On the simulator} (a) Offered and carried a load of both services for random load in Scenario $1$. (b) Performance of the RL agent in $NR$. }
  \label{fig:sc1_sim_seen_load}
\end{figure}

\noindent\textbf{Scenario $1$: evaluation of the learned policy on the simulator.} Figure \ref{fig:sc1_sim_seen_load} depicts the evaluation of the learned policy on the simulator for the random load pattern during $150$ time steps. Figure $\ref{fig:sc1_sim_seen_ls}$ gives the offered load (solid lines) and the carried load (dashed lines) of information services $1$ and $2$. We observe that the agent sometimes drops requests from service $2$ but none from service $1$. This is because a request from service $2$ is more resource-intensive than one from service $1$ and the management objective mandates maximizing the carried load. The agent's actions are thus consistent with the management objective.  
Figure $\ref{fig:sc1_sim_seen_nr}$ depicts the performance of the trained agent expressed in $NR$ for the same time period. We see that there are times when the policy is not optimal, i.e. $NR < 1$. From Table \ref{tab:anr_results} we read that the overall performance of the learned policy on the simulator for random load is $0.99$ in $ANR$. We explain the small gap to $1$ by the fact that the learned policy did not fully converge during the training period and by the probabilistic behavior of the PPO agent. 
%Overall the evaluation of this scenario suggests that the agent has been successfully trained. 

\begin{figure}[!htbp] 
  \centering
  \subfigure[Offered and carried load]{\label{fig:sc1_sim_unseen_ls}
     \includegraphics[width=\columnwidth]{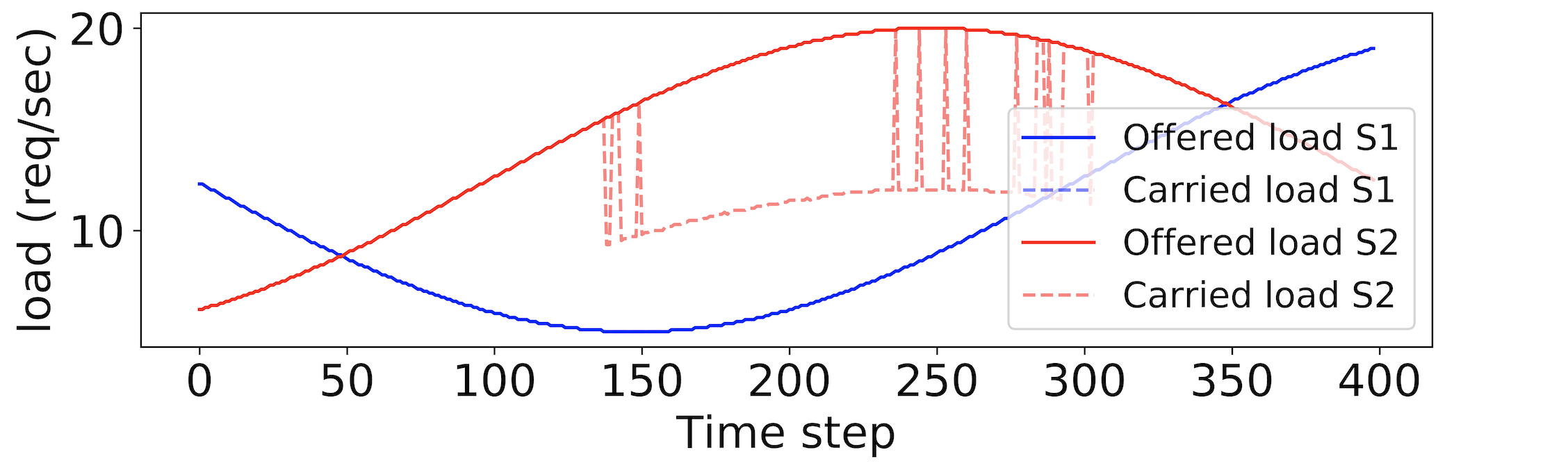}} \hspace{0.15cm}
  \subfigure[Performance of the RL agent]{\label{fig:sc1_sim_unseen_nr}
     \includegraphics[width=\columnwidth]{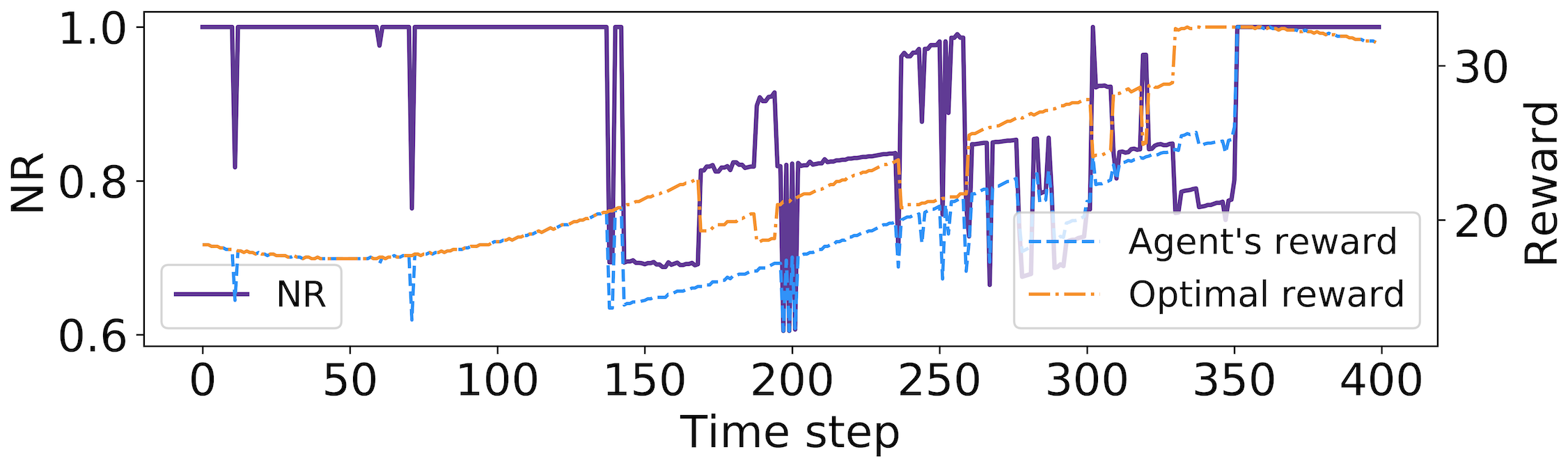}}
  \caption{\textbf{On the simulator }(a) Offered and carried load of both services for sinusoidal load in Scenario $1$. (b) Performance of the RL agent in $NR$.}
  \label{fig:sc1_sim_unseen_load}
\end{figure}

Figure $\ref{fig:sc1_sim_unseen_load}$ depicts the result of the evaluation of the agent for the sine load pattern during $400$ time steps. Figure \ref{fig:sc1_sim_unseen_ls} shows the offered (solid lines) and carried load (dashed lines) for both services, and Figure $\ref{fig:sc1_sim_unseen_nr}$ gives the performance of the agent for the same time period. We observe that, as with the random load pattern, the agent sometimes blocks requests of service $2$ when the system experiences high load, which is consistent with $MO1$. Also, we see that the agent reward can be lower than $1$ in $NR$ and can change significantly over time. We explain the occurrence of large changes with the discretized action space, which limits the  actions the agent can take. The overall performance of the agent is $0.95$ in $ANR$, which is lower than for the random load pattern. This can be explained by the fact that the sine pattern includes load values that are not seen by the agent during training. 

This evaluation suggests that the learned policy can generalize to an unseen load pattern. 

\begin{figure}[!htbp] 
  \centering
  \subfigure[Offered and carried load]{\label{fig:sc1_emu_seen_ls}
     \includegraphics[width=\columnwidth]{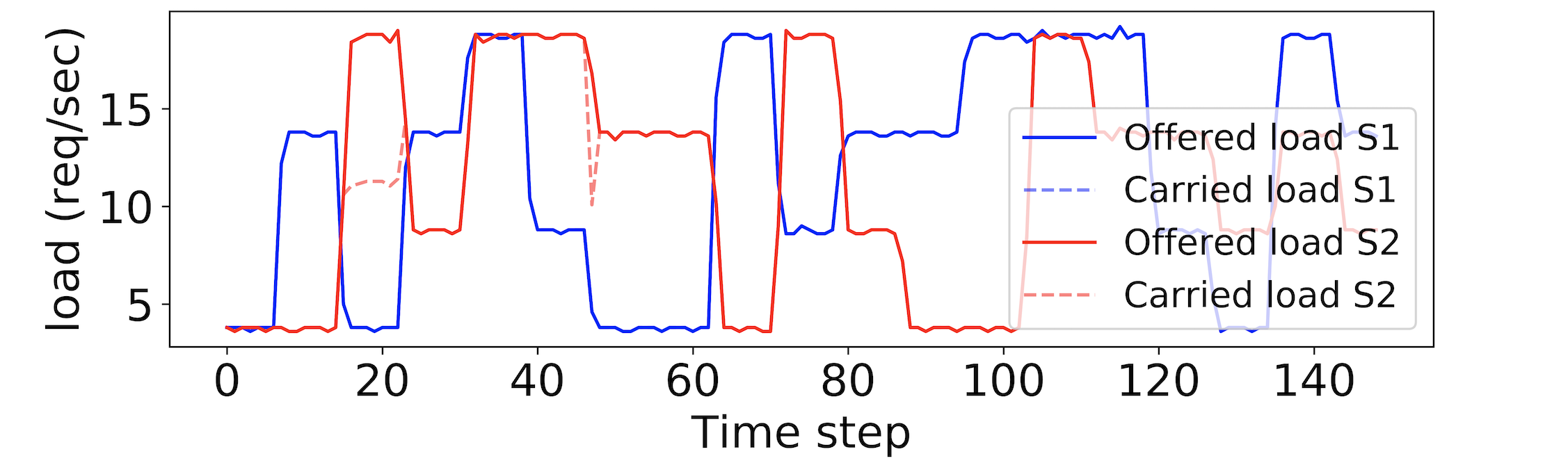}} \hspace{0.15cm}
  \subfigure[Performance of the RL agent]{\label{fig:sc1_emu_seen_nr}
     \includegraphics[width=\columnwidth]{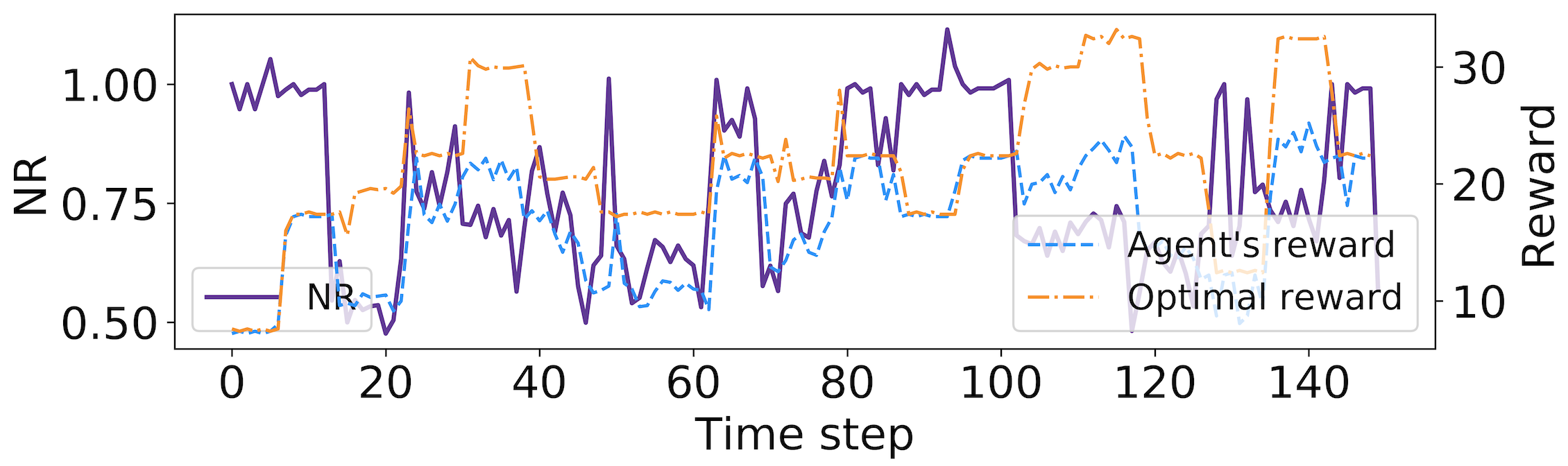}}
  \caption{\textbf{On the target system }(a) Offered and carried load of both services for random load in Scenario $1$. (b) Performance of the RL agent in $NR$.}
  \label{fig:sc1_emu_seen_load}
\end{figure}

\noindent\textbf{Scenario $1$: evaluation of the learned policy on the target system.} Figure $\ref{fig:sc1_emu_seen_load}$ shows the evaluation result of the learned policy on the target system for the random load pattern. Figure $\ref{fig:sc1_emu_seen_ls}$ gives the offered load and the carried load for information service $1$ and $2$. The load values in this experiment are very similar to those in Figure $\ref{fig:sc1_sim_seen_ls}$. The small differences are caused by the complexity of the target environment. Figure $\ref{fig:sc1_emu_seen_nr}$ illustrates the effectiveness of the learned policy on the target system. From Table \ref{tab:anr_results} we read that the overall effectiveness of the learned policy is $0.85$ in $ANR$, which is lower than $0.99$ for the same load pattern on the simulator. This gap can be explained by the inaccuracy of the system model which is about $10\%$ in $NMAE$ (if the true response time is $10$ milliseconds, the system model makes an error of $1$ millisecond on average). 

The performance of the agent of $0.85$ in $ANR$ shows the effectiveness of the learned policy on the target system.

\begin{figure}[!htbp] 
  \centering
  \subfigure[Offered and carried load]{\label{fig:sc1_emu_unseen_loads}
     \includegraphics[width=\columnwidth]{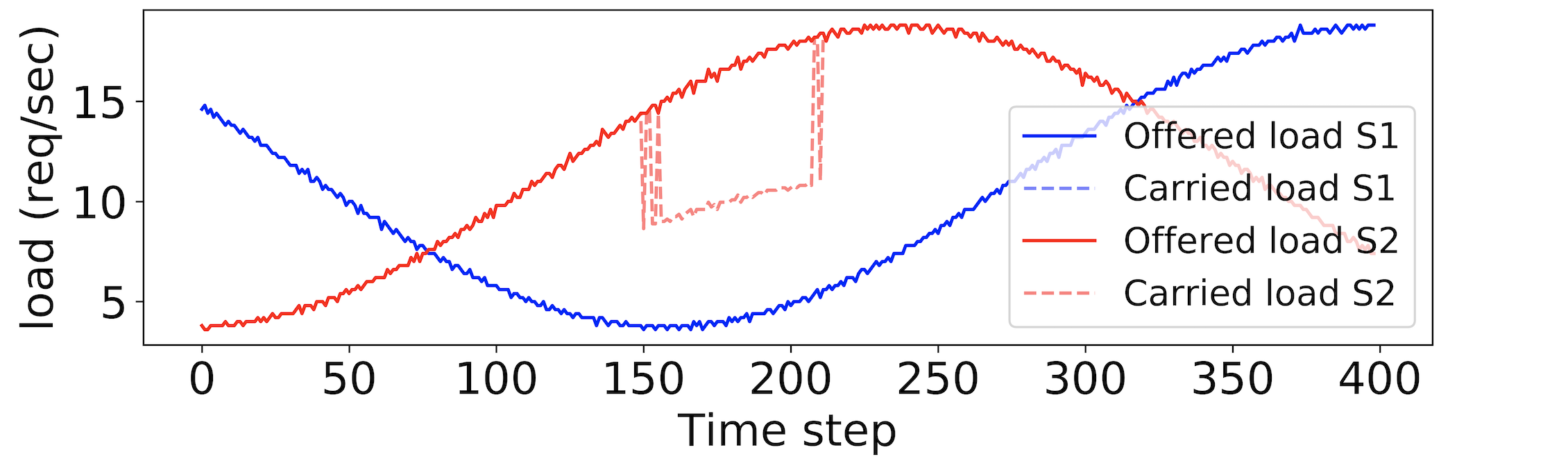}} \hspace{0.15cm}
  \subfigure[Performance of the RL agent]{\label{fig:sc1_emu_unseen_nr}
     \includegraphics[width=\columnwidth]{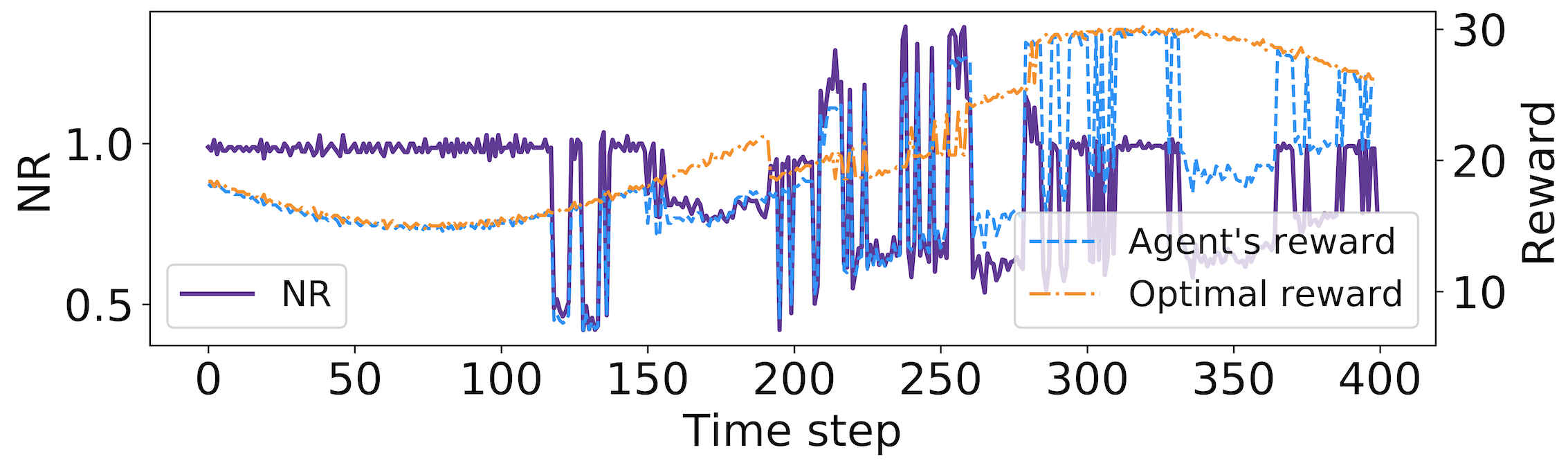}}
  \caption{\textbf{On the target system} (a) Offered and carried load of both services for sinusoidal load in Scenario $1$. (b) Performance of the RL agent in $NR$. }
  \label{fig:sc1_emu_unseen_load}
\end{figure}

Figure $\ref{fig:sc1_emu_unseen_load}$ gives the evaluation of the learned policy on the target system for the sine load pattern. Figure $\ref{fig:sc1_emu_unseen_loads}$ shows the offered load and the carried load. The figure is similar to Figure $\ref{fig:sc1_sim_unseen_ls}$; again the small differences are caused by the target environment. The performance of the learned policy can be gathered from Figure $\ref{fig:sc1_emu_unseen_nr}$. The $ANR$ of the agent for this experiment is $0.90$ (see Table \ref{tab:anr_results}). Surprisingly, this performance value is higher than that for the random load pattern on the target system although the random load pattern was used to train the agent. We believe that this anomaly is due to a combination of three factors mentioned before: the inaccuracy of the system model, the convergence gap of the RL agent, and the discretization of the action space.  

The evaluation results from Scenario $1$ demonstrate that we can train an agent that performs with high $ANR$ on the target system for an unseen load pattern.

%%%%%%%%%%%%%%%%%%%%%%%%%%%%%%%%%%%%%%%%%%%%%%%%%%%%%%%%%%%%%%%%%%%%%
%%%%%%%%%%%%%%%%%%%Evaluation results of Scenario 2%%%%%%%%%%%%%%%%%%
%%%%%%%%%%%%%%%%%%%%%%%%%%%%%%%%%%%%%%%%%%%%%%%%%%%%%%%%%%%%%%%%%%%%%
\noindent \textbf{Evaluation results of Scenario $2$.} Similar to Scenario $1$, we evaluate the learned policy for Scenario $2$ in terms of the effectiveness of the learned policy on the target system, generality, and robustness (see Table \ref{tab:anr_results}). The evaluation results of the learned policy for the random load pattern on the simulator and on the target system are $0.95$ and $0.81$ in $ANR$ respectively. These values show that the agent can learn an effective policy on the target system. The evaluation of the learned policy for the sine load pattern is $0.91$ in $ANR$ that shows to a high extent, the agent can generalize the learned policy for the unseen load pattern on the simulator. Finally, the evaluation of the learned policy for the sine load pattern on the target system is $0.80$ in $ANR$ which shows the robustness of the learned policy to the inaccuracy of the system model and the RL agent. In this scenario, as we expected the $ANR$ values on the simulator are higher than the target system, and the values of the evaluations for the random load pattern for both the simulator and target system are higher than the evaluation of the learned policy for the sine load pattern.  

\begin{figure}[!htbp] 
  \centering
  \subfigure[Offered and carried load]{\label{fig:sc2_emu_unseen_loads}
     \includegraphics[width=\columnwidth]{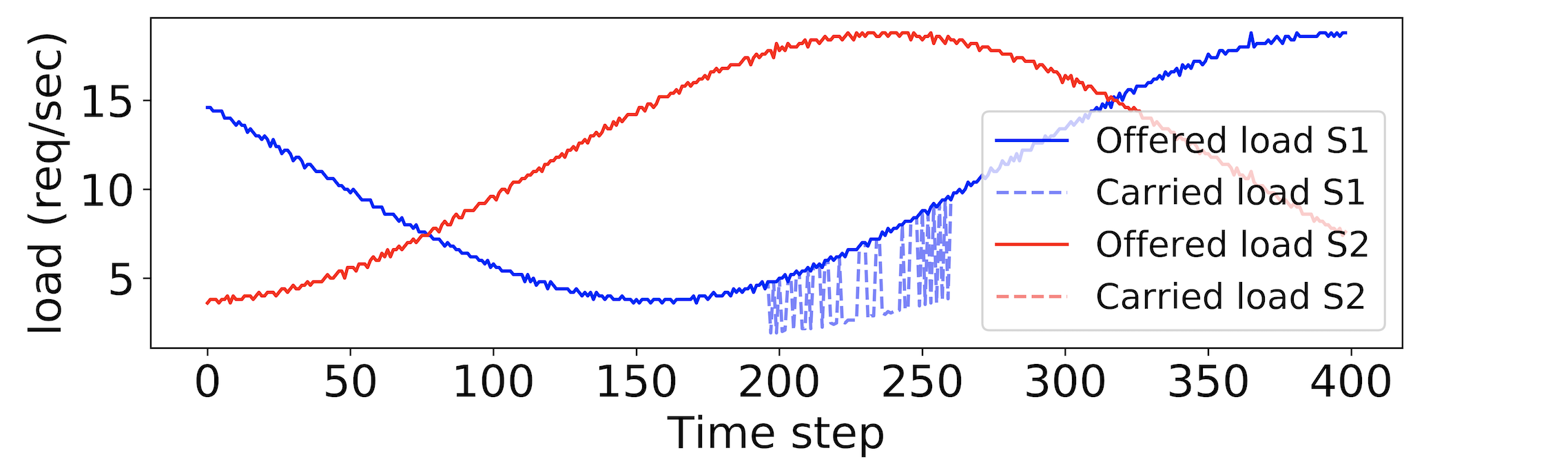}} \hspace{0.15cm}
  \subfigure[Performance of the RL agent]{\label{fig:sc2_emu_unseen_nr}
     \includegraphics[width=\columnwidth]{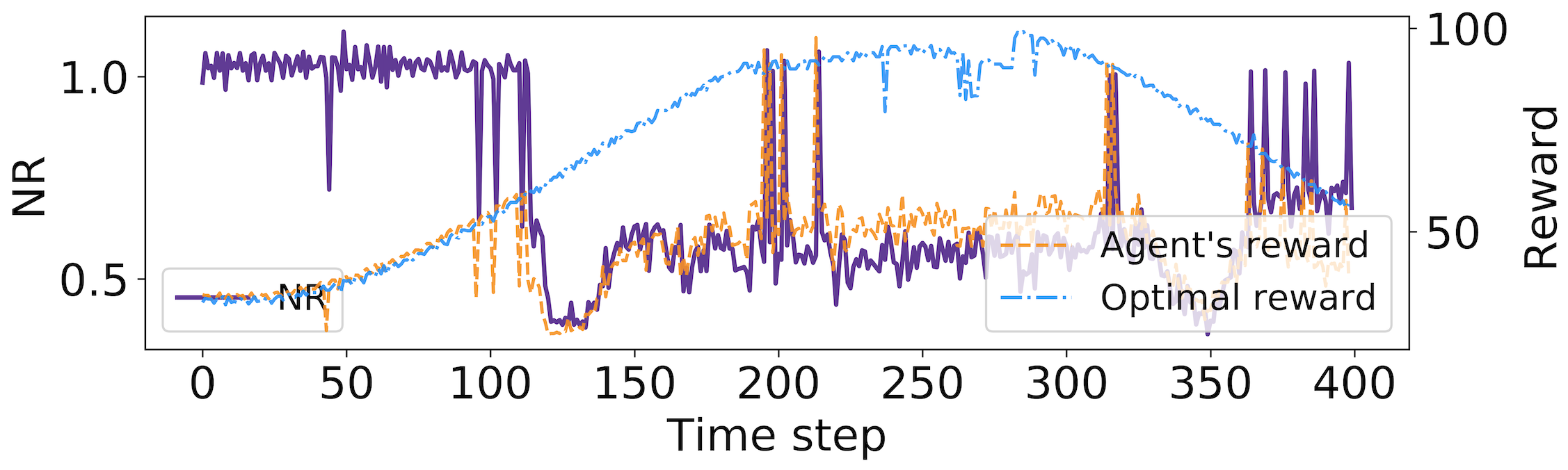}}
  \caption{\textbf{On the target system} (a) Offered and carried load of both services for sinusoidal load in Scenario $2$. (b) Performance of the RL agent in $NR$. Measurements are taken from the testbed.}
  \label{fig:sc2_emu_unseen_load}
\end{figure}

Figure $\ref{fig:sc2_emu_unseen_load}$ shows the evaluation of the learned for the sine load pattern on the target system. Figure $\ref{fig:sc2_emu_unseen_loads}$ shows the offered and carried load for this evaluation. As we can see in this figure when the offered load for service $2$ is high, the agent starts dropping requests from service $1$ (time interval of $180-250$). This behavior is consistent with the defined management objective $MO2$ for this scenario. In $MO2$, we prioritize service $2$ over service $1$, therefore, the agent drops requests from service $1$ to maximize the throughput of services with respect to the fact that service $2$ is $5$ times more important than service $1$ as well as meeting the delay constraints for both services. Comparing this figure with a similar figure in Scenario $1$, for a very similar load pattern for evaluation of the learned policy on the target system for $MO1$ (see Figure $\ref{fig:sc1_emu_unseen_load}$), we observe different behavior of the agent for dropping requests from these services. This observation shows that the agent learns the policy consistent with the defined management objective for each scenario.

%%%%%%%%%%%%%%%%%%%%%%%%%%%%%%%%%%%%%%%%%%%%%%%%%%%%%%%%%%%%%%%%%%%%%
%%%%%%%%%%%%%%%%%%%Evaluation results of Scenario 3%%%%%%%%%%%%%%%%%%
%%%%%%%%%%%%%%%%%%%%%%%%%%%%%%%%%%%%%%%%%%%%%%%%%%%%%%%%%%%%%%%%%%%%%
\noindent \textbf{Evaluation of Scanrio $3$.} Similar to previous scenarios, the evaluation results of the learned policy for the random and sine load pattern on the simulator and the target system in $ANR$, show the learned policy can perform effectively on the target system and be generalized to the unseen load values in high extent. Moreover, the learned policy can be robust to the inaccuracy of the system model and the RL agent convergence. The $ANR$ values for this scenario are presented in Table \ref{tab:anr_results}.

\begin{figure}[!htbp] 
  \centering
  \subfigure[Offered and carried load]{\label{fig:sc3_emu_unseen_loads}
     \includegraphics[width=\columnwidth]{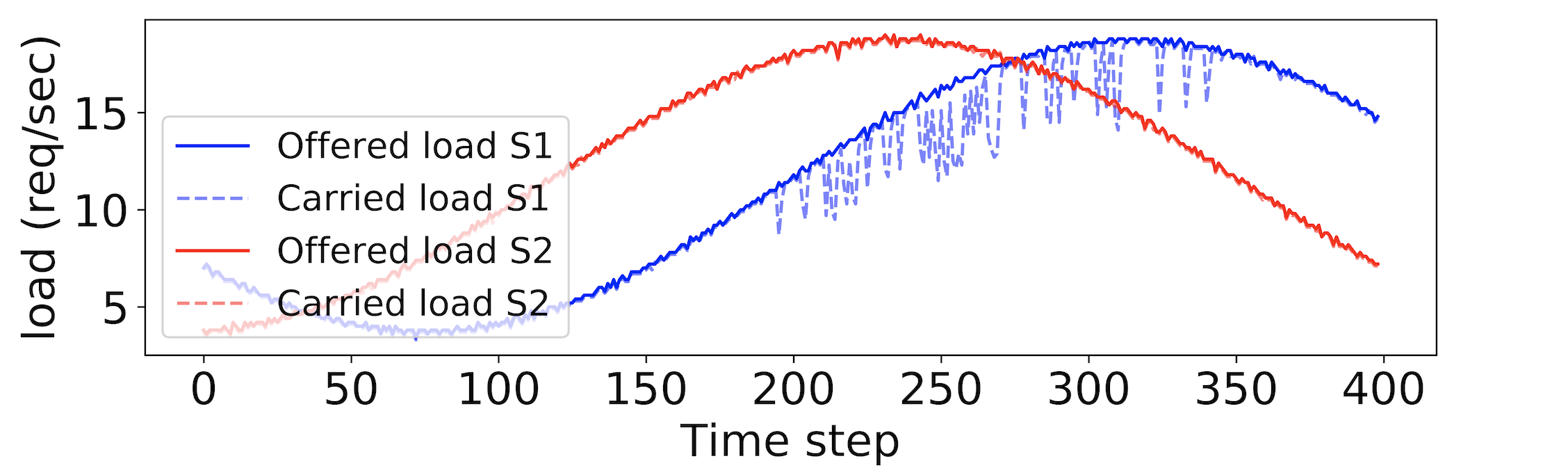}} \hspace{0.15cm}
  \subfigure[Performance of the RL agent]{\label{fig:sc3_emu_unseen_nr}
     \includegraphics[width=\columnwidth]{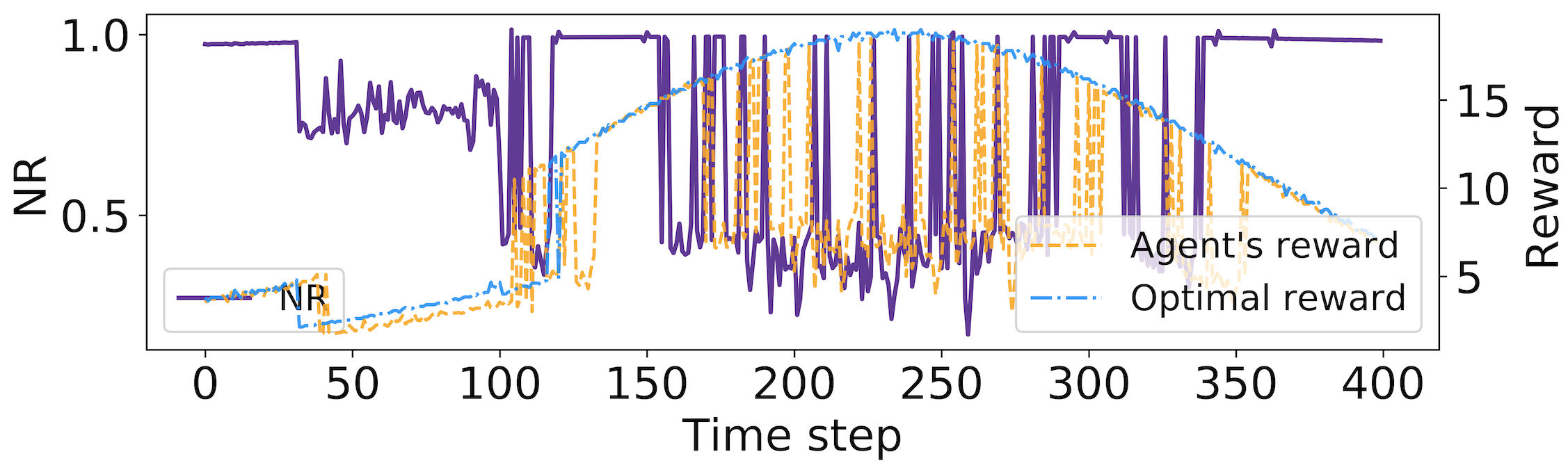}}
  \caption{\textbf{On the target system} (a) Offered and carried load of both services for sinusoidal load in Scenario $3$. (b) Performance of the RL agent in $NR$. Measurements are taken from the testbed.}
  \label{fig:sc3_emu_unseen_load}
\end{figure}

Figure $\ref{fig:sc3_emu_unseen_load}$ shows the behavior of the agent for the sine load pattern on the target system. Figure $\ref{fig:sc3_emu_unseen_loads}$ shows the offered load and carried load for both services. As we can see in this figure, the agent drops requests from service $1$ when the load values of the services increase. This behavior for the RL agent is expected since in $MO3$, we attempt to maximize the throughput of service $2$ while we prevent service $1$ from starving as well as the delays constraint for each service is met.  

%%%%%%%%%%%%%%%%%%%%%%%%%%%%%%%%%%%%%%%%%%%%%%%%%%%%%%%%%%%%%%%%%%%%%
%%%%%%%%%%%%%%%%%%%Evaluation results of Scenario 4%%%%%%%%%%%%%%%%%%
%%%%%%%%%%%%%%%%%%%%%%%%%%%%%%%%%%%%%%%%%%%%%%%%%%%%%%%%%%%%%%%%%%%%%

\noindent \textbf{Evaluation of Scenario $4$.} In this scenario, we run two services with different characteristics regarding resource consumption. One of them is information service $2$ which was run in Scenarios $1,2,\text{ and } 3$. The second service is CPU-intensive and the response time to a service request is very sensitive to the allocated CPU resource. The control functions in this scenario are routing and scaling. By scaling action, we mean changing the CPU allocation to a microservice, which is realized by vertical scaling of a Kubernetes pod. Scaling has a much higher settling time than routing or blocking, and consequently, we choose the time step for the RL agent to be $60$ seconds in this scenario instead of $5$ seconds like in Scenarios $1,2,\text{ and } 3$. 

We evaluate the learned policy for the random and the sine load pattern on the simulator and the target system in the same manner as in Scenarios $1,2,\text{ and } 3$. Similarly, the evaluation results show that the learned policy is effective on the target system, and it generalizes to an unseen load pattern. Moreover, the learned policy is sufficiently robust to the inaccuracy of the learned system model and the convergence gap of the RL agent. The evaluation results of the learned policy are summarized in Table \ref{tab:anr_results}. 

\begin{figure}[!htbp] 
  \centering
  \subfigure[Offered and carried load]{\label{fig:sc4_emu_unseen_loads}
     \includegraphics[scale=0.46]{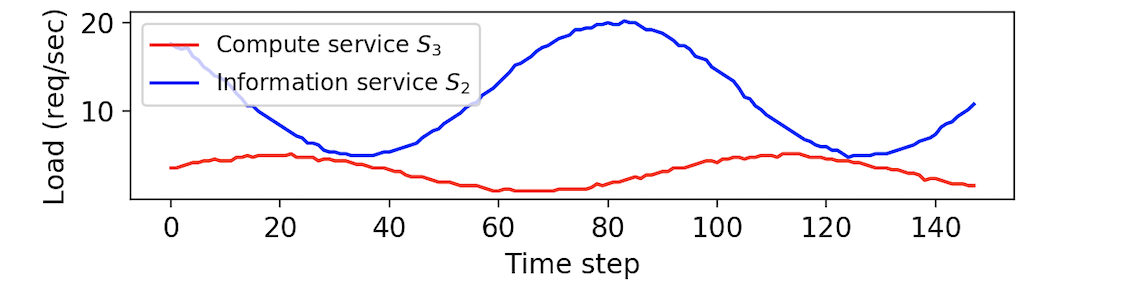}} \hspace{0.15cm}
  \subfigure[Performance of the RL agent]{\label{fig:sc4_emu_unseen_nr}
     \includegraphics[width=\columnwidth]{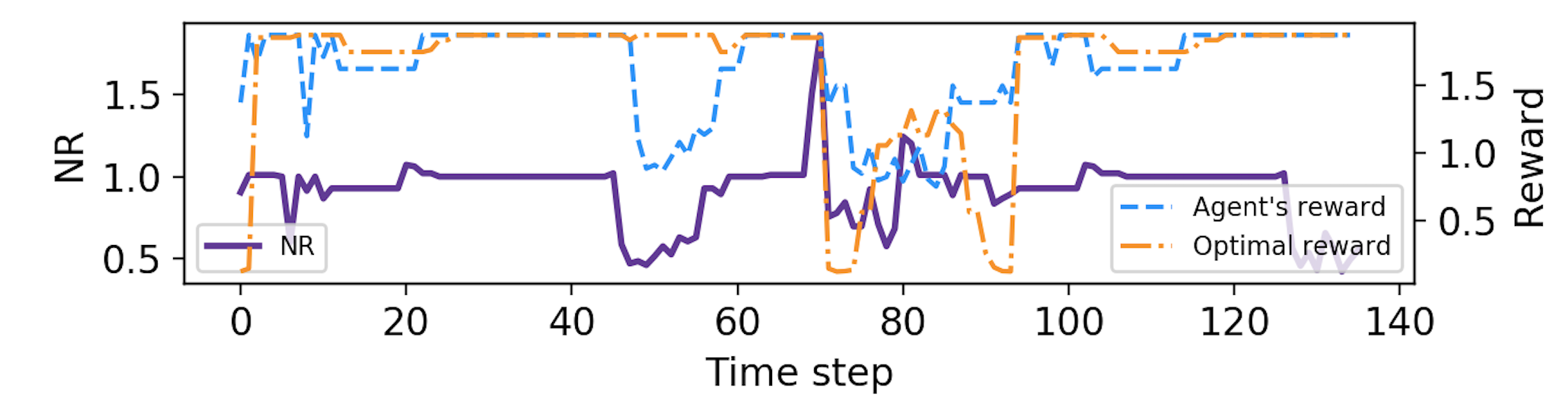}}\hspace{0.15cm}
  \subfigure[Response time of service $1$]{\label{fig:sc4_emu_unseen_response1}
     \includegraphics[width=\columnwidth]{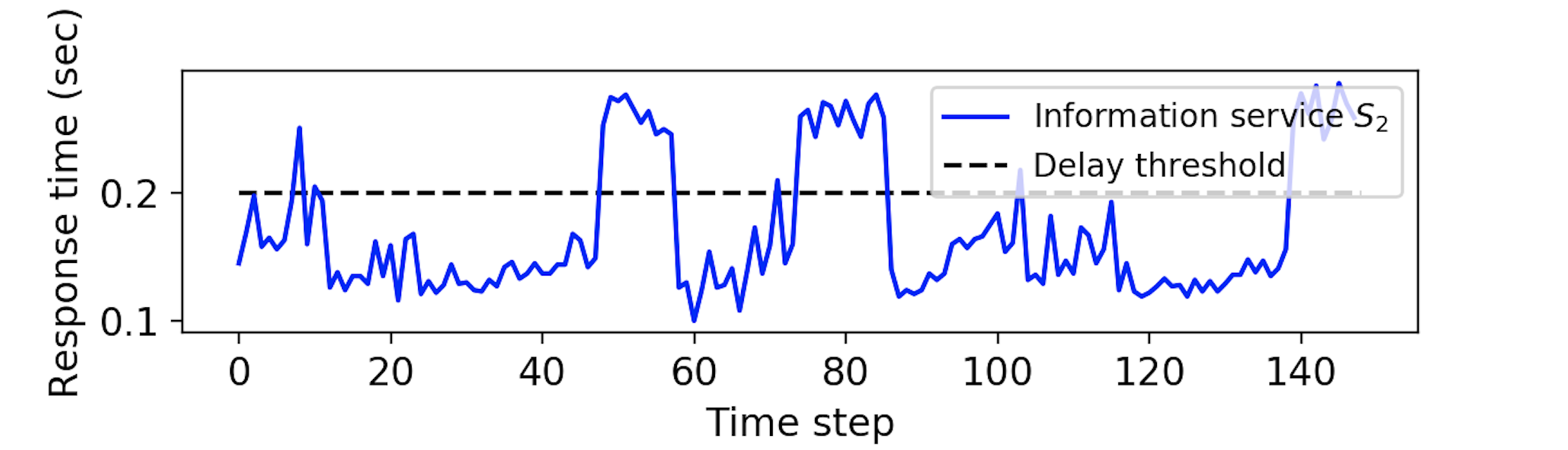}}\hspace{0.15cm}
  \subfigure[Response time of service $2$]{\label{fig:sc4_emu_unseen_response2}
     \includegraphics[width=\columnwidth]{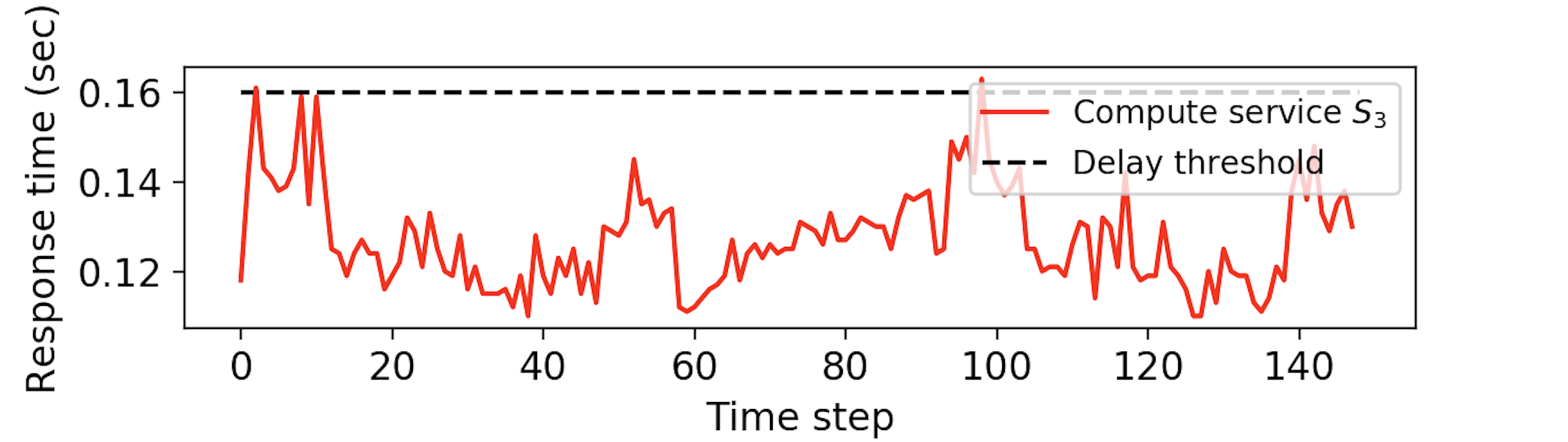}}\hspace{0.15cm}
  \subfigure[Scaling action on node $1$]{\label{fig:sc4_emu_unseen_cpu1}
     \includegraphics[width=\columnwidth]{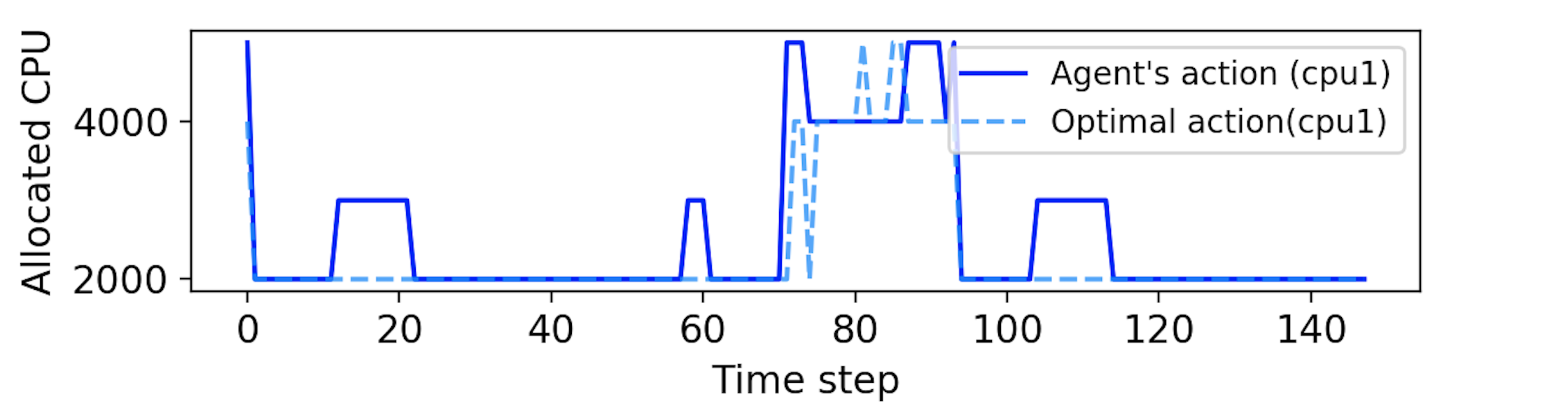}}
     \hspace{0.15cm}
  \subfigure[Scaling action on node $2$]{\label{fig:sc4_emu_unseen_cpu2}
     \includegraphics[width=\columnwidth]{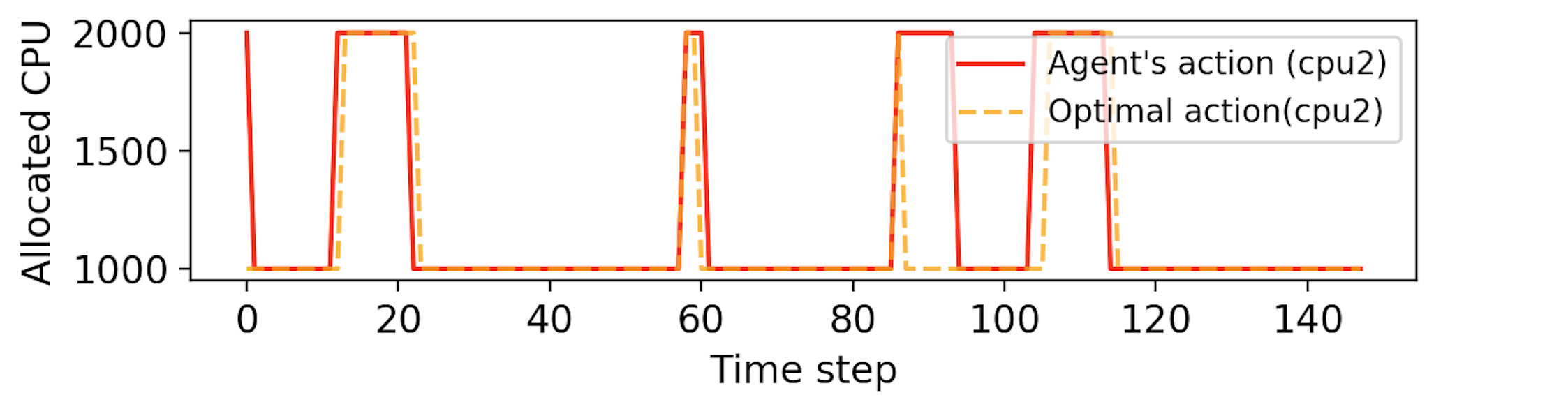}}
  \caption{\textbf{On the target system} (a) Offered and carried load of both services information service and compute service in Scenario $4$.}
  \label{fig:sc4_emu_unseen}
\end{figure}

Figure $\ref{fig:sc4_emu_unseen}$ shows the evolution of various service and performance metrics, as well as the evolution of the control settings during the experiment. Figure $\ref{fig:sc4_emu_unseen_loads}$ shows the offered and carried load for both information service $S_2$ and compute service $S_3$ following the sine load pattern. Figure $\ref{fig:sc4_emu_unseen_nr}$ shows the $NR$ values within the same time steps. Figures $\ref{fig:sc4_emu_unseen_response1}$ and $\ref{fig:sc4_emu_unseen_response2}$ show the response times of the information service and compute service. We can observe that the $NR$ values are lower than the maximum value of $1$ when the response time of service $2$ is higher than the specified threshold. Figures $\ref{fig:sc4_emu_unseen_cpu1}$ and $\ref{fig:sc4_emu_unseen_cpu2}$ show the optimal scaling actions for given states. As we can observe, the agent action follows the optimal action in most of the time intervals for both service pods. 

These results show that the agent can effectively perform on the target system for an unseen load pattern. 

\noindent\textbf{Time analysis.} We discuss the benefits and drawbacks of including a simulator in our framework. For instance, for all scenarios, training the agent on the simulator reduces the policy effectiveness by $0.11$ $ANR$ on average on unseen load (see Table \ref{tab:anr_results}). However, it significantly reduces the overall training time. For instance, for scenarios 1 to 3, this requires some $2\times 10^4$ measurement rounds, each lasting a time step of $5$ seconds, resulting in $28$ hours of monitoring time. Hyper-parameters tuning and learning effective policy for these scenarios on the testbed needs $2\,499$ hours (this time for Scenario 4 is several times larger). However, this time can be reduced to an hour on the simulator. Therefore, the training time is reduced by a factor of at least $86$. The reduction in time is achieved at the cost of reducing the policy effectiveness by $0.11$ $ANR$. 

\begin{table}[!htbp]
\centering
\caption{Performance of RL agent in all scenarios, running on the simulator and the testbed. Values are given in $ANR$. A value of $1$ means optimal performance where the management objective is met at all times.}
\label{tab:anr_results}
\begin{tabular}{|c|c|c|c|}
\hline
Scenario & Environment & Load pattern & $ANR$\\\hline
1 & simulation & random & $0.99 \pm 0.010$ \\
1 & simulation & sinusoidal & $0.95 \pm 0.014$ \\
1 & testbed & random & $0.85 \pm 0.013$ \\
1 & testbed & sinusoidal & $0.90\pm 0.012$ \\\hline
2 & simulation & random & $0.95 \pm 0.008$ \\
2 & simulation & sinusoidal & $0.91\pm 0.006$ \\ 
2 & testbed & random & $0.81\pm 0.014$\\
2 & testbed & sinusoidal & $0.80\pm 0.017$ \\\hline
3 & simulation & random & $0.99\pm 0.002$ \\
3 & simulation & sinusoidal & $0.98\pm 0.005$ \\
3 & testbed & random & $0.85\pm 0.016$ \\
3 & testbed & sinusoidal & $0.80\pm 0.015$ \\\hline
4 & simulation & random & $0.99\pm 0.005$ \\
4 & simulation & sinusoidal & $0.98\pm 0.010$ \\
4 & testbed & random & $0.84 \pm 0.042$ \\
4 & testbed & sinusoidal & $0.91 \pm 0.037$ \\\hline
\end{tabular}
\end{table} 

\section{Related work}
\label{sec:related_work}

The work described in this paper falls into several broad areas of traditional and recent research in resource management, namely \textit{self-adaptive systems} (surveys \cite{morandini2008towards}\cite{weyns2012survey}\cite{krupitzer2015survey}), \textit{autonomic computing and networking} (e.g., \cite{kephart2003vision}\cite{parashar2004autonomic}), \textit{self-driving systems} (e.g., \cite{schneider2020self}) and \textit{intent-based networking} (see e.g., \cite{rafiq2020intent}\cite{campanella2019intent}). The solutions our framework gives rise to are self-adaptive since the agent policy dynamically adapts the system to a changing environment; they are autonomous because this adaptation is performed without human intervention; they follow self-driving principles in the sense that they navigate the system towards a specific objective; and they are intent-based because they implement an end-to-end management objective.  

Our work specifically focuses on adaptive resource management for cloud services using reinforcement learning. There are several survey papers covering this topic, for instance, \cite{gari2021reinforcement} and \cite{cardellini2019self}. (The main difference between them is that \cite{gari2021reinforcement} reviews work that uses virtual machines (VMs) while \cite{cardellini2019self} discusses papers that use containers instead). The objectives considered in the cited papers are similar to our work. They contain variables like resource utilization, quality of service, and cost. The control functions used to meet the objectives include scaling, scheduling, and migration. A common problem these investigations face stems from the fact that reinforcement learning often requires long training times, which we discuss in detail below.

\begin{table*}
\centering
\caption{Important works on adaptive resource management for microservice-based applications using reinforcement learning}
\label{tab:related_work}
\scalebox{1}{
\begin{tabular}{|l|c|c|c|c|c|}
\hline
\multirow{3}{*}{Paper}& & &\multirow{2}{*}{RL System} & \multirow{2}{*}{Training} &  \multirow{2}{*}{Evaluation} \\
& Performance objective & Resource controls &  \multirow{2}{*}{model}&\multirow{2}{*}{environment} &  \multirow{2}{*}{environment} \\
& & &&&\\\hline

\multirow{4}{*}{\cite{xu2022coscal}, Xu, 2022} &\multirow{3}{*}{Resource utilization}& \multirow{3}{*}{Vertical and horizontal} & \multirow{3}{*}{Learned from} & Initialize policy & \multirow{4}{*}{Target system}\\
&\multirow{3}{*}{Response time}&\multirow{3}{*}{scaling}&\multirow{3}{*}{external traces}&from external traces&\\
&  & &  & Update policy on & \\
&&&&Target system&\\\hline

\multirow{3}{*}{\cite{schneider2021self}, Schneider, 2021} & \multirow{2}{*}{Response time and}  & \multirow{3}{*}{Routing and placement} & Learned from a & \multirow{3}{*}{Simulation} & \multirow{3}{*}{Simulation}\\
 & \multirow{2}{*}{throughput} &  & digital twin &  & \\
  & &  & simulator &  & \\\hline

\multirow{2}{*}{\cite{garg2021heuristic}, Grag 2021} & Deployment cost and  &\multirow{2}{*}{Placement and migration}& \multirow{2}{*}{-} & \multirow{2}{*}{Simulation} & \multirow{2}{*}{Simulation} \\
 & response time  &  &  &  &  \\\hline

\multirow{2}{*}{\cite{rossi2020self}, Rossi 2020} & Resource cost and & Vertical and horizontal & Learned from  & \multirow{2}{*}{Simulation} & \multirow{2}{*}{Simulation}\\
 & response time& scaling & target system &  & \\\hline

\multirow{2}{*}{\cite{qiu2020firm}, Qiu 2020} & Reource utilization and & Vertical and horizontal& Learned from  & \multirow{2}{*}{Target system} & \multirow{2}{*}{Target system} \\
 & response time & scaling& target system &  &  \\\hline

\multirow{3}{*}{\cite{yang2019miras}, Yang 2019} & Finished tasks & \multirow{3}{*}{Vertical scaling} & \multirow{2}{*}{Learned from} & \multirow{3}{*}{Simulation} & \multirow{3}{*}{Target system} \\
 & which is correlated &  & \multirow{2}{*}{target system} &  &  \\
 & to the response time &  &  &  &  \\\hline

\multirow{4}{*}{This paper} & General form including & \multirow{2}{*}{Routing} & &  \multirow{4}{*}{Simulation} &  \multirow{4}{*}{Target system}\\
&end-to-end response time& \multirow{2}{*}{Admission control}& Learned from&&\\
&throughput& \multirow{2}{*}{Vertical scaling} &target system&&\\
&cost and utility&&&&\\\hline
\end{tabular}
}
\end{table*} 

%%%
%1. top-down -> paragraph 3
%2. Generalit -> paragraph 3
%3. Proactive -> paragraph 4
%4. Applying on the real system and efficiency because of simulator -> paragraph 5

Resource allocation is investigated in many papers as the problem of designing a single control function, such as horizontal scaling \cite{park2021graf} \cite{delande2021horizontal} \cite{rajib2022lightweight} \cite{yu2020microscaler}\cite{faticanti2020throughput} \cite{qiu2020firm} \cite{yang2019miras} \cite{xu2022coscal} \cite{rossi2019horizontal} \cite{ayimba2021sqlr} \cite{li2023task} \cite{jiang2021resource}, application-layer routing \cite{ruuskanen2022dynamical}, or placement and migration of the microservice instances. \cite{garg2021heuristic}\cite{wang2019delay}\cite{schneider2021self}\cite{lin2021client}. When designing such a function, the authors associate a performance objective, for example meeting a delay constraint for an application while minimizing the CPU allocation \cite{park2021graf}\cite{delande2021horizontal}\cite{faticanti2020throughput}, or meeting SLO constraints while minimizing  energy consumption  \cite{hou2020ant}\cite{samanta2020dyme}. 

Note that we take a different approach with our framework. We start with an end-to-end performance objective and then direct the available resource control functions to jointly meet the objective. The RL agent takes a combination of actions, whereby one type of action corresponds to one resource management function. In the scenarios studied in this paper, the management objective can be achieved through any combination of three functions, request routing, blocking, and scaling. (If desired, additional control functions like placement or horizontal scaling can be organically added to our framework.) Therefore, our approach can be described as top-down, and it is more general than virtually all related works we are aware of, which create bottom-up solutions that are use-case specific.  

In the literature resource management problems are generally formulated as optimization problems with constraints. Among the works studying adaptive resource management on microservice-based applications, we find these problems formulated using mathematical frameworks like control theory \cite{baarzi2021showar}, decision theory \cite{yang2019miras}, or scheduling theory \cite{xu2020self}. Approaches to solve these optimization problems include reinforcement learning \cite{qiu2020firm}\cite{yang2019miras}\cite{wang2019delay}\cite{xu2022coscal}\cite{rossi2019horizontal}\cite{garg2021heuristic}, dynamic programming \cite{wu2023towards}, contextual bandit\cite{delande2021horizontal}, and Bayesian optimization \cite{yu2020microscaler}. Reinforcement learning has become a popular approach in recent research, because it allows controlling complex systems with little knowledge about their structure and internal workings and is therefore well suited for cloud-based applications. 

The main challenge of applying reinforcement learning to resource management is the long training time of the agent, which originates from the large number of interactions of the agent with the environment that are needed to optimally solve a resource allocation problem. Several strategies are pursued in the literature to address this issue, often in combination. The first strategy is called model-based reinforcement learning \cite{wang2019delay}\cite{garg2021heuristic}\cite{rossi2020self}. It includes creating a mathematical model of the environment, for instance using queuing theory \cite{ruuskanen2022dynamical} or caching methods \cite{bao2019performance}. Alternatively, it uses supervised learning to obtain a system model \cite{stadler2017learning}\cite{park2021graf}. This is the approach we are pursuing in this paper. The second strategy is to control only a small part of the target system, for instance a single microservice, in a single control action \cite{qiu2020firm}. This significantly reduces the complexity of the control problem and requires a much smaller number of interactions of the agent with the target system. This strategy is often used with reactive schemes that resolve performance bottlenecks. A third approach consists of discretizing the action space and sometimes the state space as well,  thus restricting the number of possible actions and states to small numbers. This limits the number of possible state-action pairs that need to be considered by the training algorithm \cite{joseph2019fuzzy}. In our work, we limit the number of possible control actions for this reason (see Section \ref{sec:problem_formulation_and_approach}). The downside of this approach is that the management objective is achieved with respect to the discretized control and state spaces, which can result in suboptimal control. The fourth strategy to reduce training time entails training the agent on a simulator instead of on the target system. Note that this method does not reduce the number of interactions of the agent with the environment but takes advantage of the fact that the interaction time on a simulator is typically orders of magnitude smaller than on the target system. For this reason, we have included a simulator in our framework.      

Orchestration platforms like Kubernetes include automated resource management functions. For example, Kubernetes Horizontal Pod Autoscaling (HPA) \cite{K8autoscaler} performs horizontal scaling based on thresholds that the user can set. The system then performs scaling actions to keep the CPU utilization below the given threshold. This autoscaling function can be included in our framework and expand the control capabilities we have implemented in our testbed. However, note that this function alone is not an alternative to our framework, since it only uses CPU consumption thresholds to trigger autoscaling \cite{qiu2020firm} and cannot in a meaningful way implement complex management objectives like those discussed in this paper \cite{casalicchio2017auto}\cite{toka2020adaptive}\cite{nguyen2020horizontal}.

% One of the industrial solutions for auto-scaling is Kubernetes Horizontal Pod Autoscaling (HPA) \cite{K8autoscaler}. However, this solution has limits such as only using the CPU consumption rules to trigger autoscaling \cite{qiu2020firm}, lack of possibility to define a complex MO for an end-to-end metric to trigger the autoscaling and sensitivity to the monitoring metrics (relative metrics versus absolute) \cite{casalicchio2017auto}\cite{toka2020adaptive}\cite{nguyen2020horizontal}. 

Important related works on adaptive resource management for microservice-based applications using reinforcement learning are listed in Table \ref{tab:related_work}. All use reinforcement learning to find the optimal control policy. All works (except ours) address a very specific use case and do not develop a general approach like we introduce in this paper. \cite{xu2022coscal} is of interest because it uses external traces from an operational environment to initialize the policy. \cite{garg2021heuristic} does not consider a system model but instead solves an optimal placing problem using reinforcement learning. 
The authors of \cite{qiu2020firm} train an RL agent on the target system in an online fashion. They perform resource management in a reactive manner as follows. Once an SLO violation is detected, a microservice component is identified as a possible cause. Then, the allocation of resources to this particular microservice is adjusted using reinforcement learning.  
The authors in \cite{yang2019miras} follow the same approach as we do by training the agent in a simulation environment and evaluating the learned policy on the target system. They study resource management for scientific workflows, whereby resources are allocated to microservices that process workflow tasks in order to minimize the average processing time for a task.

\section{Conclusions and future work}
\label{sec:conclusion}
We demonstrated how end-to-end management objectives under varying load can be met through periodic control actions performed by a RL agent. By computing, near-optimal control policies on a simulator, effective control on a testbed can be achieved for several scenarios with applications on a service mesh, which suggests that the approach and framework we propose are practical.

After a thorough review of the related literature, we find that our proposed framework is unique in the following way. First, it advocates a top-down approach whereby the management objectives are defined first and then mapped onto the available control actions. Second, management objectives for several services can be defined and achieved simultaneously on an infrastructure. Third, several control actions can be executed simultaneously in a single time step of the RL agent. Fourth, our framework is general in the sense that many classes of management objectives as well as a long list of control actions can be supported. 

Like other recent studies, we learn a system model from traces of the target system. Unlike these works, however, we also learn the operating region which constrains the possible action of the agent and we consider the different settling times of the various control actions, which facilitates meeting the management objectives. In addition, as we showed in Scenarios 1 to 3, we can support different management objectives with the same system model, a useful property that we have not seen discussed anywhere else in the literature.      

The numerical results show that by using the simulator in our framework, we can speed up the RL agent training by a factor of at least 100 in our setup. 

In our future work, we want to validate the framework for practical microservice-based applications. Most practical applications include a larger number of microservices than those we used for the scenarios in this paper. To perform resource management for such applications, we expect the action space to have higher dimensionality and therefore the number of possible actions an agent can take will increase significantly. One approach we plan to follow is to sequentialize certain control actions which are now performed in parallel. This means we reformulate the control problem from a single-stage problem utilized in this work to a multistage problem and we study the trade-off between the number of interactions required to train the agent on the simulator with the number of the control steps required to meet the management objective on the target system. 

Second, we will study online policy adaptation. In our current approach, the system model and control policy are learned sequentially and in an offline fashion. In an operational environment, however, the policy must adapt at runtime in response to system failures or changes to the system configuration. We plan to extend our framework towards online learning of the system model with periodic updates. These updates are read by the simulator, which continuously produces updated control policies and sends them to the RL agent on the testbed.

\section*{Acknowledgements}
The authors are grateful to Andreas Johnsson, Farnaz Moradi, Jalil Taghia, Xiaou Lan, and Hannes Larsson with Ericsson Research for the fruitful discussion around this work. The authors especially thank Kim Hammar at KTH for his help in developing part of the RL agent and for constructive comments on a draft of this paper. The authors thank Xiaoxuan Wang at KTH for her helpful comments on a draft of this paper. This research has been partially supported by the Swedish Governmental Agency for Innovation Systems, VINNOVA, through project ANIARA. 

% \balance
\bibliographystyle{IEEEtran}
\bibliography{tnsm}

% \vspace*{-3\baselineskip}
\vskip -1\baselineskip plus -1fil

\begin{IEEEbiography}[{\includegraphics[width=1.0in,height=1.25in,clip,keepaspectratio]{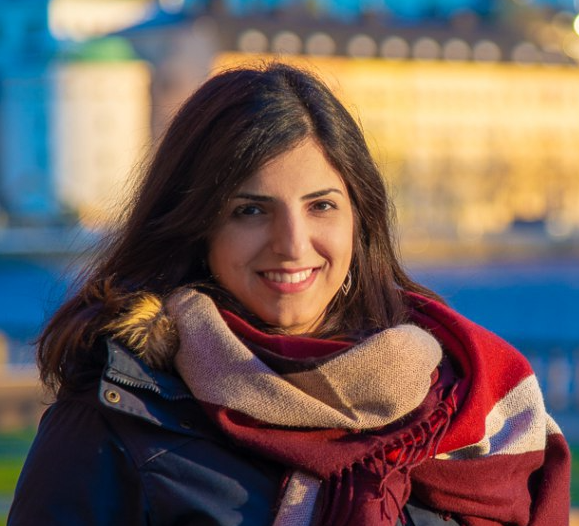}}]{Forough Shahab Samani}
is a Ph.D. student at KTH Royal Institute of Technology in Stockholm, Sweden. She holds an M.Sc degree in computer engineering from the Sharif University of Technology in Tehran, Iran. Before starting her Ph.D., she worked in the R\&D group at ISC (Informatics Services Corporation) in Tehran, Iran, for four years. 
\end{IEEEbiography}
% \vfill
\vskip -1\baselineskip plus -1fil
% \vspace*{-2\baselineskip}
\begin{IEEEbiography}[{\includegraphics[width=1.0in,height=1.25in,clip,keepaspectratio]{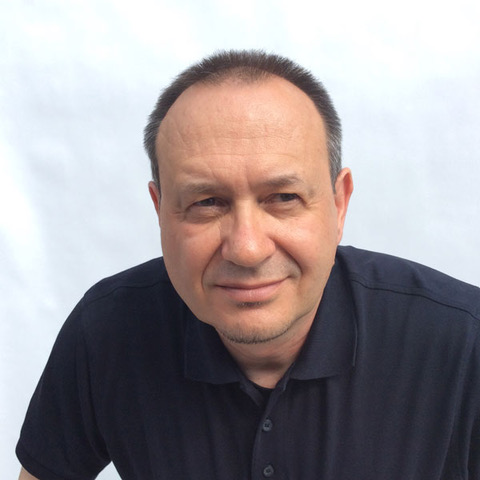}}]{Rolf Stadler}
is a professor at KTH Royal Institute of Technology in Stockholm, Sweden, and head of the Division of Network and Systems Engineering. He holds an M.Sc. degree in mathematics and a Ph.D. in computer science from the University of Zurich. Before joining KTH in 2001, he held positions at the IBM Zurich Research Laboratory, Columbia University, and ETH Z\"urich. His group has made contributions to real-time monitoring, resource management, and self-management for large-scale networked systems. His current interests include advanced monitoring techniques and data-driven methods for network engineering and management. Rolf Stadler has been Editor-in-Chief of IEEE Transactions on Network and Service Management (TNSM) from 2014-2017.
\end{IEEEbiography}

\end{document}